\documentclass[notitlepage,a4paper,aps,prd,onecolumn,superscriptaddress,nofootinbib,groupedaddress]{revtex4}

\usepackage{amsmath,comment}
\usepackage{amsfonts,color}
\usepackage{amsmath}
\usepackage{amsthm}
\usepackage{graphicx} 
\usepackage{pdfpages}
\usepackage{amsmath}
\usepackage{empheq}
\usepackage{amsfonts,color}
\usepackage{amssymb,float}
\usepackage{physics}
\usepackage{booktabs,multirow}
\usepackage{titlesec}

\titleformat{\paragraph}
{\normalfont\normalsize\bfseries}{\theparagraph}{1em}{}
\titlespacing*{\paragraph}
{0pt}{3.25ex plus 1ex minus .2ex}{1.5ex plus .2ex}

\usepackage[none]{hyphenat}

\usepackage{amsmath}
\usepackage[utf8]{inputenc}
\allowdisplaybreaks
\usepackage{color}
\usepackage{hyperref}
\hypersetup{
    colorlinks=true,
    linkcolor=blue,
    filecolor=magenta,      
   citecolor=blue
}

\usepackage{enumitem}

\usepackage{tikz}
\usetikzlibrary{shapes.geometric}
\usetikzlibrary{arrows.meta,arrows}

\begin{document}
\title{A gravitational spin-orbit interaction in Poincaré gauge theory}

\author{Sebastian Bahamonde}
\email{sbahamondebeltran@gmail.com}
\affiliation{Kavli Institute for the Physics and Mathematics of the Universe (WPI), The University of Tokyo Institutes
for Advanced Study (UTIAS), The University of Tokyo, Kashiwa, Chiba 277-8583, Japan.}
\affiliation{Cosmology, Gravity, and Astroparticle Physics Group, Center for Theoretical Physics of the Universe,
Institute for Basic Science (IBS), Daejeon, 34126, Korea.}

\author{Jorge Gigante Valcarcel}
\email{jorgevalcarcel@ibs.re.kr}
\affiliation{Center for Geometry and Physics, Institute for Basic Science (IBS), Pohang 37673, Korea.}

\begin{abstract}

We show a gravitational spin-orbit interaction that can potentially modify the space-time geometry naturally emerges in the framework of Poincaré gauge theory. For this purpose, we derive the field equations of a particular model with cubic order invariants and demonstrate the existence of analytical solutions which display an interaction between the intrinsic and extrinsic angular momentum parameters in the gravitational action, in analogy to the spin-orbit interaction arising from atomic and nuclear systems. Due to the highly nonlinear character of the field equations under stationary and axisymmetric conditions, we focus on a degenerate case which simplifies their complexity, at the cost of constraining the geometry to the Kerr space-time. Thereby, our results indicate more general solutions with a spin-orbit interaction beyond the Kerr space-time are expected to arise in the nondegenerate models of Poincaré gauge theory.

\end{abstract}

\maketitle

\section{Introduction}

The spin-orbit interaction (SOI) is a fundamental coupling that plays a critical role in shaping the energy levels of atoms, the structure of atomic nuclei and the electronic properties of materials~\cite{Thomas:1926dy,Thomas:1927yu,Mayer:1948zz,Mayer:1949pd,Haxel:1949fjd,vzutic2004spintronics}. However, despite of its widely recognised importance in accurately describing microscopic phenomena governed by the electromagnetic and nuclear fields, its application in macroscopic gravitational systems has not yet been explored to the same extent.

Certainly, our current understanding of gravity is based on the physical correspondence of the space-time curvature with the energy and momentum properties of matter, as featured in the celebrated Einstein's field equations of General Relativity (GR)~\cite{Wald:1984rg}. This geometrical picture leads to a large number of solutions describing a wide range of astrophysical and cosmological phenomena~\cite{Stephani:2003tm,Griffiths:2009dfa}. Among these results, the Kerr solution remarkably describes the space-time geometry of a rotating massive black hole~\cite{Kerr:1963ud}, while further generalisations also exist due to the gravitational effect of the energy-momentum tensor of the electromagnetic and Yang-Mills fields~\cite{Newman:1965tw,Newman:1965my,Volkov:1997qb,Volkov:1998cc,Kleihaus:2000kg}.

On the other hand, the extension of GR towards a post-Riemannian geometry with torsion enables the study on the physical implications of the intrinsic spin of matter in the space-time~\cite{Kibble:1961ba,sciama1962analogy,Sciama:1964wt}. The gravitational field characterised by curvature and torsion is then described by a gauge field associated with the external rotations and translations of the Poincaré group~\cite{Hehl:1976kj,Obukhov:1987tz,Blagojevic:2013xpa,ponomarev2017gauge,Obukhov:2022khx}. Thereby, on top of the energy-momentum tensor of matter, a gravitating spin tensor acting as a source of torsion arises in this approach.

In this context, the dynamical behaviour of torsion can in fact modify the metric structure of the black hole solutions of GR~\cite{Baekler:1981lkh,mccrea1987kerr,Obukhov:1987tz,Cembranos:2016gdt,Cembranos:2017pcs,Obukhov:2019fti,Obukhov:2020hlp,Bahamonde:2021qjk}. Of particular interest is the study of rotating black holes under stationary and axisymmetric conditions, since the interaction provided by the respective intrinsic and extrinsic angular momentum parameters may have a relevant impact in their geometry. However, the computational intractability underlying from the high nonlinearity of the field equations of Poincaré Gauge (PG) theory with curvature and torsion represents a serious obstacle to carry out these studies, which has eventually restricted the search of rotating black hole solutions to the conventional Kerr-de Sitter and Kerr-Newman space-times~\cite{mccrea1987kerr,Obukhov:2019fti,Bahamonde:2021qjk} (see also~\cite{Bakler:1988nq,Garcia:1998jw,Bahamonde:2022meb} for natural generalisations with electromagnetic and NUT charges, cosmological constant and acceleration parameter), modulo sign differences in the Lagrangian coefficients appearing in the solutions.

Following these lines, in this letter we address the search of rotating black hole solutions in cubic PG theory, which has been recently considered to eliminate ghostly instabilities from the vector and axial modes of the torsion field described by quadratic PG theory~\cite{Bahamonde:2024sqo}. Given the high mathematical complexity of the field equations of cubic PG theory in stationary and axisymmetric space-times, our main goal is to find analytical solutions which could display a gravitational SOI between the intrinsic and extrinsic angular momentum parameters in the gravitational action, regardless the cubic PG model under consideration restricts the form of the metric tensor to the conventional Kerr solution. Indeed, we will show this can be naturally obtained in a degenerate model devoid of kinetic terms for the axial mode of torsion, where the respective field equations acquire a relatively simpler form in comparison to the nondegenerate models of the theory and it is easier to find such solutions. Thereby, the present work is aimed to introduce the possible effects of the gravitational SOI in the space-time geometry, which can be consistently described in the framework of PG theory.

This paper is organised as follows. First, in Sec.~\ref{sec:model} we introduce the cubic PG model with curvature and torsion considered in our study, for which we derive the field equations following the gauge principles of PG theory. For the sake of simplicity in the presentation, we provide the specific form of the field equations in Appendix~\ref{appe1}, whereas a specific list of the Lagrangian coefficients fixed by our stability conditions can be found in Appendix~\ref{appe2}. Then, in Sec.~\ref{subsec:axialA} we impose stationarity and axial symmetry, in order to extend previous studies on static and spherically symmetric black hole solutions of cubic PG theory under these symmetries~\cite{Bahamonde:2024sqo}. Unfortunately, the underlying system of equations becomes in general intractable, even under the slow rotation approximation. Thus, we are led to restrict our attention to a degenerate model, for which the field equations acquire a relatively simpler form that allows us to follow a systematic procedure to solve them, at the cost of constraining the geometry to the slowly rotating Kerr space-time. Thereby, in Sec.~\ref{sec:solutions} we present analytical solutions displaying a gravitational SOI for this model. In fact, it turns out that the cubic order invariants constructed from the torsion and curvature tensors of PG theory are crucial to  provide such solutions with interaction terms between the intrinsic and extrinsic angular momentum parameters in the gravitational action, which represents another distinctive feature of cubic PG theory. Finally, we present the conclusions in Sec.~\ref{sec:conclusions}.

We work in natural units $c=G=\hbar=1$ and consider the metric signature $(+,-,-,-)$. In addition, we use a tilde accent to denote quantities defined from the general affine connection that includes torsion and nonmetricity, in order to distinguish them from their unaccented counterparts defined from the Levi-Civita connection. On the other hand, Latin and Greek indices run from $0$ to $3$, referring to anholonomic and coordinate bases, respectively.

\section{PG model with cubic order invariants}\label{sec:model}

A gauge approach to gravity arises naturally when the unitary irreducible representations of relativistic particles labeled by their spin and mass are linked to the space-time geometry. Then, a gauge connection valued in the Lie algebra of the Poincar\'{e} group can be introduced to describe the gravitational interaction as a gauge field of the external rotations and translations \cite{Hehl:1976kj,Obukhov:1987tz,Blagojevic:2013xpa,ponomarev2017gauge,Obukhov:2022khx}:
\begin{equation}
    A_{\mu}=e^{a}{}_{\mu}P_{a}+\omega^{a b}{}_{\mu}J_{a b}\,,
\end{equation}
where $e^{a}{}_{\mu}$ and $\omega^{a b}{}_{\mu}$ respectively represent the tetrad field and the spin connection, satisfying the following relations with the metric tensor $g_{\mu\nu}$ and the affine connection $\tilde{\Gamma}^{\lambda}{}_{\rho\mu}$ of a Riemann-Cartan (RC) space-time:
\begin{align}
    g_{\mu \nu}&=e^{a}{}_{\mu}\,e^{b}{}_{\nu}\,\eta_{a b}\,,\\
    \label{anholonomic_connection}
    \omega^{a b}{}_{\mu}&=e^{a}{}_{\lambda}\,e^{b\rho}\,\tilde{\Gamma}^{\lambda}{}_{\rho \mu}+e^{a}{}_{\lambda}\,\partial_{\mu}\,e^{b\lambda}\,,
\end{align}
with $\eta_{a b}$ the local Minkowski metric. On the other hand, $P_{a}$ and $J_{a b}$ correspond to the generators of the Poincar\'{e} group, fulfilling commutation relations:
\begin{align}
    \left[P_{a},P_{b}\right]&=0\,,\\
    \left[P_{a},J_{bc}\right]&=i\,\eta_{a[b}\,P_{c]}\,,\\
    \left[J_{ab},J_{cd}\right]&=\frac{i}{2}\,\left(\eta_{ad}\,J_{bc}+\eta_{cb}\,J_{ad}-\eta_{db}\,J_{ac}-\eta_{ac}\,J_{bd}\right).
\end{align}

Thus, the corresponding field strength tensor derived from the gauge connection includes translational and rotational parts: 
\begin{align}
    F^{a}{}_{\mu\nu}&=\partial_{\mu}e^{a}{}_{\nu}-\partial_{\nu}e^{a}{}_{\mu}+\omega^{a}{}_{b\mu}\,e^{b}{}_{\nu}-\omega^{a}{}_{b\nu}\,e^{b}{}_{\mu}\,,\\
    F^{ab}{}_{\mu\nu}&=\partial_{\mu}\omega^{a b}{}_{\nu}-\partial_{\nu}\omega^{a b}{}_{\mu}+\omega^{a}{}_{c\mu}\,\omega^{c b}{}_{\nu}-\omega^{a}{}_{c\nu}\,\omega^{c b}{}_{\mu}\,,
\end{align}
which are indeed related to the torsion and curvature tensors
\begin{align}
    T^{\lambda}\,_{\mu \nu}&=2\tilde{\Gamma}^{\lambda}\,_{[\mu \nu]}\,,\\
    \tilde{R}^{\lambda}\,_{\rho \mu \nu}&=\partial_{\mu}\tilde{\Gamma}^{\lambda}\,_{\rho \nu}-\partial_{\nu}\tilde{\Gamma}^{\lambda}\,_{\rho \mu}+\tilde{\Gamma}^{\lambda}\,_{\sigma \mu}\tilde{\Gamma}^{\sigma}\,_{\rho \nu}-\tilde{\Gamma}^{\lambda}\,_{\sigma \nu}\tilde{\Gamma}^{\sigma}\,_{\rho \mu}\,,
\end{align}
as follows:
\begin{align}
    F^{a}\,_{\mu\nu}&=e^{a}\,_{\lambda}T^{\lambda}\,_{\nu\mu}\,,
\\
    F^{a b}{}_{\mu\nu}&=e^{a}\,_{\lambda}e^{b\rho}\tilde{R}^{\lambda}\,_{\rho\mu\nu}\,.
\end{align}

In order to endow these quantities with dynamics, the structure of the Poincaré group entails a large number of curvature and torsion invariants in the gravitational action. In this sense, different restrictions imposed on the
Lagrangian coefficients lead to a broad class of PG models, for which an extensive number of fundamental properties and phenomenological implications may arise. Thus, one of the most relevant aspects to be considered in this construction concerns stability, which has been recently analysed in the presence of cubic order invariants to provide a safer context for the propagation of torsion around general backgrounds\footnote{Note that the Riemannian Einstein-Hilbert term naturally arises from the general action of PG theory, in virtue of the identity
\begin{equation}
    \tilde{R}=R-2\nabla_{\mu}T^{\nu \mu}\,_{\nu}+\frac{1}{4}T_{\lambda \mu \nu}T^{\lambda \mu \nu}+\frac{1}{2}T_{\lambda \mu \nu}T^{\mu \lambda \nu}-T^{\lambda}\,_{\lambda\nu}T^{\mu}\,_{\mu}\,^{\nu}\,.
\end{equation}}~\cite{Bahamonde:2024sqo}:
\begin{align}
    S=&\;\frac{1}{16\pi}\int\bigl[-\,R-\frac{1}{2}\left(2c_{1}+c_{2}\right)\tilde{R}_{\lambda\rho\mu\nu}\tilde{R}^{\mu\nu\lambda\rho}+c_{1}\tilde{R}_{\lambda\rho\mu\nu}\tilde{R}^{\lambda\rho\mu\nu}+c_{2}\tilde{R}_{\lambda\rho\mu\nu}\tilde{R}^{\lambda\mu\rho\nu}+d_{1}\tilde{R}_{\mu\nu}\bigl(\tilde{R}^{\mu\nu}-\tilde{R}^{\nu\mu}\bigr)
\Bigr.\,
\nonumber\\
\Bigl.
&+\bar{h}_{1}^{} \tilde{R}_{\rho \tau \gamma \mu } T_{\alpha }{}^{\gamma \mu \
} T^{\alpha \rho \tau }+ \bar{h}_{2}^{} \tilde{R}_{\rho \gamma \tau \mu } 
T_{\alpha }{}^{\gamma \mu } T^{\alpha \rho \tau } + \bar{h}_{3}^{} \tilde{R}_{\alpha \tau \gamma \mu } T^{\alpha \rho \tau } T_{\rho }{}^{
\gamma \mu } + \bar{h}_{4}^{} \tilde{R}_{\alpha \gamma \tau \mu } T^{\alpha 
\rho \tau } T_{\rho }{}^{\gamma \mu } + \bar{h}_{5}^{} \tilde{R}_{\tau \gamma \alpha \mu } T^{\alpha \rho \tau } T_{\rho }{}^{\gamma \mu }\nonumber\\
&+ \bar{h}_{6}^{} \tilde{R}_{\gamma \mu \alpha \tau } T^{\alpha \rho \tau } T_{\rho }{}^{\gamma \mu }+ \bar{h}_{7}^{} \tilde{R}_{\rho \tau \gamma \mu } T^{\tau \gamma \mu } T^{\lambda\rho }{}_{\lambda} + \bar{h}_{8}^{} \tilde{R}_{\rho \gamma \tau \mu } T^{\tau \gamma \mu } T^{\lambda\rho }{}_{\lambda} + \bar{h}_{9}^{} \tilde{R}_{\tau \gamma \rho \mu } T^{\tau \gamma \mu } T^{\lambda\rho }{}_{\lambda} + \bar{h}_{10}^{} \tilde{R}_{\gamma \mu \rho \tau } T^{\tau \gamma \mu } T^{\lambda\rho }{}_{\lambda}\nonumber\\
&+ \bar{h}_{11}^{} \tilde{R}_{\alpha \tau \gamma \mu } T^{\alpha \rho \tau } T^{\gamma }{}_{\rho }{}^{\mu }+\bar{h}_{12}^{} \tilde{R}_{\alpha \gamma \tau \mu } T^{\alpha \rho \tau } T^{\gamma }{}_{\rho }{}^{\mu } + \bar{h}_{13}^{} \tilde{R}_{\alpha \mu \tau \gamma } T^{\alpha \rho \tau } T^{\gamma }{}_{\rho }{}^{\mu } + \bar{h}_{14}^{} \tilde{R}_{\tau \mu \alpha \gamma } T^{\alpha \rho \tau } T^{\gamma }{}_{\rho }{}^{\mu }+\bar{h}_{15}^{} \tilde{R}^{\alpha \rho } T_{\alpha }{}^{\tau \gamma } T_{\rho \tau \gamma }\nonumber\\
& + \bar{h}_{16}^{} \tilde{R}^{\alpha \rho } T_{\rho }{}^{\tau \gamma } T_{\tau \alpha \gamma }+ \bar{h}_{17}^{} \tilde{R}^{\alpha \rho } T_{\alpha }{}^{\tau \gamma } T_{\tau \rho \gamma } + \bar{h}_{18}^{} \tilde{R}^{\alpha \rho } T_{\tau \rho \gamma } T^{\tau }{}_{\alpha }{}^{\gamma } + \bar{h}_{19}^{} \tilde{R}^{\alpha \rho } T^{\tau }{}_{\alpha }{}^{\gamma } T_{\gamma \rho \tau }+ \bar{h}_{20}^{} \tilde{R}^{\alpha \rho } T^{\sigma}{}_{\alpha\sigma}T^{\lambda}{}_{\rho\lambda}\nonumber\\
&+ \bar{h}_{21}^{} \tilde{R}^{\alpha \rho } T_{\alpha \rho }{}^{\beta } T^{\lambda}{}_{\beta\lambda}+ \bar{h}_{22}^{} \tilde{R}^{\alpha \rho } T_{\rho \alpha }{}^{\beta} T^{\lambda}{}_{\beta\lambda}+ \bar{h}_{23}^{} \tilde{R}^{\alpha \rho } T^{\beta}{}_{\alpha \rho } T^{\lambda}{}_{\beta\lambda} +\bar{h}_{24}^{} \tilde{R} T_{\alpha \rho \tau } T^{\alpha \rho \tau }+ 
\bar{h}_{25}^{} \tilde{R} T^{\alpha \rho \tau } T_{\rho \alpha \tau } + 
\bar{h}_{26}^{} \tilde{R} T^{\lambda}{}_{\rho\lambda}T^{\sigma\rho }{}_{\sigma}\nonumber\\
&+\frac{1}{2}m^{2}_{T}T^{\lambda}{}_{\mu\lambda} T^{\sigma\mu}{}_{\sigma}-3m^{2}_{S}T_{[\lambda\mu\nu]}T^{[\lambda\mu\nu]}
+\frac{1}{3}m^{2}_{t}\left(T_{\lambda\mu\nu}T^{\lambda\mu\nu}+T_{\lambda\mu\nu}T^{\mu\lambda\nu}-T^{\lambda}{}_{\mu\lambda} T^{\sigma\mu}{}_{\sigma}\right)\bigr]\sqrt{-\,g}\,d^{4}x\,,\label{cubicPGmodel_fulltorsion}
\end{align}
where
\begin{equation}
    \tilde{R}_{\mu\nu}=\tilde{R}^{\lambda}\,_{\mu \lambda \nu}\,, \quad \tilde{R}=g^{\mu\nu}\tilde{R}_{\mu\nu}\,.
\end{equation}

Then, the field equations are derived from the cubic PG action by performing variations with respect to the tetrad field and the spin connection, which according to the principle of least action must satisfy:
\begin{equation}
    \delta S=\frac{1}{16 \pi}\int{\left(e_{a\mu}E^{\mu\nu}\delta e^{a}\,_{\nu}+e_{a\lambda}e_{b \mu}E^{\lambda\mu\nu}\delta \omega^{a b}{}_{\nu}\right)\sqrt{-g}\,d^4x}=0\,,\end{equation}
where $E^{\mu\nu}$ and $E^{\lambda\mu\nu}$ are tensor quantities depending on curvature and torsion that are defined in Appendix~\ref{appe1}.

A post-Riemannian expansion of the gravitational action~\eqref{cubicPGmodel_fulltorsion} in terms of the irreducible modes of the torsion tensor
\begin{align}\label{Tdec_1}
T_{\mu}&=T^{\nu}\,_{\mu\nu}\,,\\
S_{\mu}&=\varepsilon_{\mu\lambda\rho\nu}T^{\lambda\rho\nu}\,,\\
t_{\lambda\mu\nu}&=T_{\lambda\mu\nu}-\frac{2}{3}g_{\lambda[\nu}T_{\mu]}-\frac{1}{6}\,\varepsilon_{\lambda\rho\mu\nu}S^{\rho}\,,\label{Tdec3}
\end{align}
allows the ghostly instabilities present in the vector and axial sectors of the theory to be identified and removed by a proper choice of the Lagrangian coefficients. Thus, this procedure sets one of the coefficients associated with the quadratic curvature invariants and four of the coefficients of the cubic invariants, so that the resulting action  can be expressed as~\cite{Bahamonde:2024sqo}:
\begin{align}
    S=&\;\frac{1}{16\pi}\int\bigl[-\,R-2c_{1}\tilde{R}_{\lambda\rho\mu\nu}\tilde{R}^{\mu\nu\lambda\rho}+c_{1}\tilde{R}_{\lambda\rho\mu\nu}\tilde{R}^{\lambda\rho\mu\nu}+2c_{1}\tilde{R}_{\lambda\rho\mu\nu}\tilde{R}^{\lambda\mu\rho\nu}+d_{1}\tilde{R}_{\mu\nu}\bigl(\tilde{R}^{\mu\nu}-\tilde{R}^{\nu\mu}\bigr)
\Bigr.\,
\nonumber\\
\Bigl.
&+\frac{1}{2}h_{1}\bigl(2\tilde{R}_{\mu\nu}T^{\mu}T^{\nu}-\tilde{R}T_{\mu}T^{\mu}\bigr)-\frac{1}{6}\bigl(c_{1}+6h_{13}\bigr)\tilde{R}_{\mu\nu}S^{\mu}S^{\nu}+\frac{1}{2}h_{13}\tilde{R}S_{\mu}S^{\mu}+h_5 \tilde{R}_{\lambda\rho\mu\nu}t_{\sigma}{}^{\lambda\rho}t^{\sigma\mu\nu}\nonumber\\
&+h_6\tilde{R}_{\lambda\rho\mu\nu}t_{\sigma}{}^{\lambda\mu}t^{\sigma\rho\nu}+h_7\tilde{R}_{\lambda\rho\mu\nu}t^{\lambda\rho}{}_{\sigma}t^{\sigma\mu\nu}+h_{8}{} \tilde{R}_{\lambda\rho\mu\nu}t^{\lambda\mu}{}_{\sigma}t^{\sigma\rho\nu}+h_{9}{}\tilde{R}_{\lambda\rho\mu\nu}t^{\lambda\mu}{}_{\sigma}t^{\rho\nu\sigma}+h_{10}{}\tilde{R}_{\lambda\rho}t_{\mu\nu}{}^{\lambda}t^{\rho\mu\nu}\nonumber\\
&+h_{11}{}\tilde{R}_{\lambda\rho}t_{\mu\nu}{}^{\lambda}t^{\mu\nu\rho}+h_{12}{}\tilde{R}t_{\lambda\rho\mu}t^{\lambda\rho\mu}+h_{13}\bigl(\varepsilon^{\lambda\rho\mu\nu}\tilde{R}_{\lambda\rho\mu\nu}T_{\sigma}S^{\sigma}-2\varepsilon_{\nu}{}^{\lambda\rho\sigma}\tilde{R}_{\lambda\rho\mu\sigma}T^{\mu}S^{\nu}+4\varepsilon^{\lambda\rho\mu\nu}\tilde{R}_{\lambda\rho}T_{\mu}S_{\nu}\bigr)\nonumber\\
&+h_{16}{}\tilde{R}_{\lambda\rho\mu\nu}T^{\nu}t^{\lambda\rho\mu}+h_{17}{} \tilde{R}_{\lambda\rho\mu\nu}T^{\rho}t^{\lambda\mu\nu}+h_{18}{}\tilde{R}_{\lambda\rho}T_{\mu}t^{\mu\lambda\rho}+h_{19}{} \tilde{R}_{\lambda\rho}T_{\mu}t^{\lambda\rho\mu}+h_{20}{}\varepsilon_{\alpha\rho\mu\nu}\tilde{R}_{\tau}{}^{\rho\mu\nu}S^{\gamma}t^{\alpha\tau}{}_{\gamma}\nonumber\\
&+h_{21}{}\varepsilon_{\alpha\rho\mu\nu}\tilde{R}_{\tau}{}^{\rho\mu\nu}S^{\gamma}t_{\gamma}{}^{\alpha\tau}+h_{22}{}\varepsilon_{\alpha\rho}{}^{\mu\nu}\tilde{R}^{\rho}{}_{\mu\tau\nu}S^{\gamma}t_{\gamma}{}^{\alpha\tau}+h_{23}{}\varepsilon_{\alpha\rho}{}^{\mu\nu}\tilde{R}_{\gamma\mu\tau\nu}S^{\alpha}t^{\gamma\rho\tau}+h_{24}{}\varepsilon_{\alpha\rho}{}^{\mu\nu}\tilde{R}_{\gamma\mu\tau\nu} S^{\alpha}t^{\rho\tau\gamma}\nonumber\\
&+h_{25}{}\varepsilon_{\alpha\rho\tau\mu}\tilde{R}^{\mu}{}_{\gamma}S^{\alpha}t^{\rho\tau\gamma}+h_{26}{}\varepsilon_{\lambda\rho\mu\nu} \tilde{R}^{\lambda\rho}S_{\sigma}t^{\sigma\mu\nu}+\frac{1}{2}m^{2}_{T}T_{\mu} T^{\mu}+\frac{1}{2}m^{2}_{S}S_{\mu}S^{\mu}
+\frac{1}{2}m^{2}_{t}t_{\lambda\mu\nu}t^{\lambda\mu\nu}\bigr]\sqrt{-\,g}\,d^{4}x\,,\label{cubicPGmodel_irreducible_torsion}
\end{align}
where the specific values of the aforementioned coefficients and the reparametrisation used in this expression can be found in Appendix~\ref{appe2}.

It is worthwhile to stress that the variational equations derived from the action~\eqref{cubicPGmodel_irreducible_torsion} do not satisfy in general the Birkhoff's theorem in a static and spherically symmetric space-time, but imposing further restrictions on the Lagrangian coefficients it is possible to find Reissner-Nordström-like solutions with dynamical torsion~\cite{Bahamonde:2024sqo}. Given the following line element in the static and spherically symmetric space-time
\begin{equation}\label{sph_metric}
    ds^{2}=\Psi(r)dt^{2} -\frac{dr^{2}}{\Psi(r)}-r^{2}d\vartheta^{2}-r^{2}\sin^{2}\vartheta \, d\varphi^{2}\,,
\end{equation}
and the rotated frame
\begin{align}
\vartheta^{\hat{0}}&=\frac{1}{2}\left[ \left(\Psi(r)+1\right)\,dt+\left(1-\frac{1}{\Psi(r)}\right)\,dr \right],\\
\vartheta^{\hat{1}}&=\frac{1}{2}\left[\left(\Psi(r)-1\right)\,dt+\left(1+\frac{1}{\Psi(r)}\right)\,dr \right],\\
\vartheta^{\hat{2}}&=r\,d \vartheta \,,\\
\vartheta^{\hat{3}}&=r\sin\vartheta \, d \varphi \,,
\end{align}
the general solution then arises for the Lagrangian coefficients
\begin{align}
  c_1&=-\,\frac{\left(2 N_1+N_2\right)}{2\left(4 N_1+N_2\right)}d_1 \,,\quad h_1=h_{13}=h_{16}=h_{17}=h_{18}=h_{19}=  0\,,\label{Lag_coeff1}\\
  h_{8}&=\frac{1}{5}\left[9 d_1+10 h_{7}+9 h_{25}-4\left(5 h_{5}+h_{6}\right)+\frac{8 N_1 \left(4 d_1+9 h_{25}-9 h_{6}\right)}{2 N_1+5 N_2}-\frac{10N_{1}d_{1}}{4 N_1+N_2}\right],\\
  h_{9}&=-\,h_{10}=\frac{1}{10}\left[3 d_1+20 h_{7}+3 h_{25}-8\left(5 h_{5}+h_{6}\right)\right]+\frac{8 N_1 \left(4 d_1+9 h_{25}-9 h_{6}\right)}{5 \left(2 N_1+5 N_2\right)}+\frac{N_{1}d_{1}}{4 N_1+N_2}\,,\\
 h_{11}&=2 h_{7}-4 h_{5}+\frac{9\left(2 N_1+N_2\right)\left[\left(2N_{1}+N_{2}\right)d_{1}+\left(4N_{1}+N_{2}\right)h_{25}\right]}{2\left(4 N_1+N_2\right)\left(2 N_1+5 N_2\right)}-\frac{\left(10 N_1+7 N_2\right)h_{6}}{2 N_1+5 N_2}\,,\\
  h_{12}&=-\,\frac{1}{24} \left[3\left(d_{1}+4h_{7}+h_{25}-8 h_{5}-2h_{6}\right)+\frac{8 N_1 \left(4 d_1+9 h_{25}-9 h_6\right)}{2 N_1+5 N_2}+\frac{2N_{1}d_{1}}{4 N_1+N_2}\right],\\
  h_{22}&=\frac{1}{4} \left[4 h_{21}-2 d_1-8h_{20}-5 h_{25}-4 h_{26}+\frac{3 N_1 (2 d_1+3 h_{25})}{2N_1-N_2}+\frac{N_{1}h_{25}}{2 N_1+N_2}\right],\\
  h_{23}&=-\,2\left(d_1+2 h_{20}+h_{25}\right)+\frac{2 N_1 \left(2d_{1}+3h_{25}\right)}{2 N_1-N_2}+\frac{2N_{1}d_{1}}{4N_{1}+N_{2}}\,, \quad h_{24}=\frac{1}{2}\left(h_{23}+h_{25}\right),\\
  m_{t}^{2}&=-\,\frac{6w\left(2N_1-N_2\right)\left[d_1 \left(2N_1+N_2\right)+\left(h_{25}-h_{6}\right)\left(4N_1+N_2\right)\right]}{\left(4N_1+N_2\right)\left(2N_1+5N_2\right)}\,, \quad m^{2}_{T}=m^{2}_{S}=0\,,\label{Lag_coeff24}
\end{align}
and includes Coulomb-like axial and tensor modes of torsion with a spin charge as\footnote{Note that the torsion tensor of the complete solution is decomposed into two contributions as $T^{\lambda}{}_{\mu\nu} = \mathring{T}^{\lambda}{}_{\mu\nu}+\bar{T}^{\lambda}{}_{\mu\nu}$, where $\bar{T}^{\lambda}{}_{\mu\nu}$ introduces the spin charge that for general values of the parameters $N_{1}$ and $N_{2}$ modifies the space-time geometry, and $\mathring{T}^{\lambda}{}_{\mu\nu}$ represents the background torsion that does not include any explicit dependence on this charge (see Expressions (70)-(71) in~\cite{Bahamonde:2024sqo} for the specific form of the torsion modes of this contribution, including its dependence on the factor $w$ associated with the mass terms of torsion).}:
\begin{align}
    \bar{S}_{a} &= \frac{2\left(N_{2}-2N_{1}\right)\kappa_{\rm s}}{r}\left(1, -1,0,0\right),\\
    \bar{t}^{a}{}_{b c} &= \frac{\left(N_{1}+N_{2}\right)\kappa_{\rm s}}{3r}\left(
\begin{array}{cccccc}
0 & 0 & 0 & 0 & 0 & 2 \\
0 & 0 & 0 & 0 & 0 & 2 \\
0 & 0 & -1 & 0 & 1 & 0 \\
0 & 1 & 0 & -1 & 0 & 0
\end{array} \right)\,,
\end{align}
where the columns in the matrix representation refer to antisymmetric indices in the order (01, 02, 03, 12, 13, 23). In addition, the metric function reads
\begin{equation}
    \Psi(r)=1-\frac{2m}{r}
    +\left(2N_{1}-N_{2}\right) \left(N_{1}+N_{2}\right)\left[\frac{2N_{1}+N_{2}}{4N_{1}+N_{2}}d_{1}+2h_{25}\right]\frac{\kappa_{\rm s}^{2}}{3r^{2}}\,.\label{formPsi}
\end{equation}
Therefore, the solution describes a Reissner-Nordström-like black hole for general values of the parameters $N_1$ and $N_2$, while for the specific cases where $N_2 = -\,N_1$ or $N_2 = 2N_1$ the contribution of the spin charge in the metric tensor is cancelled out. In particular, the case $N_2 = -\,N_1$ reduces the gravitational action~\eqref{cubicPGmodel_irreducible_torsion} to a degenerate model devoid of kinetic terms for the axial mode of torsion, which anticipates a reduction in the overall complexity of the field equations in the stationary and axisymmetric space-time, in comparison with the general case.

\section{Stationary and axisymmetric space-times in PG theory}

\subsection{Invariance conditions}\label{subsec:axialA}

The geometry of stationary and axisymmetric space-times is characterised by two Killing vectors $\partial_{t}$ and $\partial_{\varphi}$, which respectively generate time translations and rotations around a symmetry axis. The latter defines a regular two-dimensional timelike surface of fixed points where it vanishes and provides a metric structure which is invariant under the action of the rotation group SO(2) \cite{Stephani:2003tm,Ortin:2015hya}. For a stationary and axisymmetric space-time satisfying circularity conditions, it is possible to write the corresponding line element in a specific gauge depending on a nontrivial off-diagonal component $g_{t\varphi}$ alone, since the two Killing vectors $\partial_{t}$ and $\partial_{\varphi}$ are not mutually orthogonal \cite{hartle1967variational}:
\begin{equation}\label{axi_line}
    ds^2=\Psi_1(r,\vartheta)\,dt^2-\frac{dr^2}{\Psi_2(r,\vartheta)}-r^2\Psi_3(r,\vartheta)\left[ d\vartheta^2+\sin^2\vartheta\bigl(d\varphi-\Psi_4(r,\vartheta)dt\bigr)^2\right],
\end{equation}
where $\{\Psi_{i}\}_{i=1}^{4}$ are four arbitrary functions depending on $r$ and $\vartheta$.

This geometrical invariance can be extended to torsion, in such a way that the corresponding RC curvature tensor is also preserved under the same symmetries, which sets the general form of the stationary and axisymmetric torsion tensor as
\begin{equation}
    T^{\lambda}{}_{\mu\nu} = T^{\lambda}{}_{\mu\nu} \left(r,\vartheta\right).
\end{equation}
Thus, the number of degrees of freedom (dof) included in the torsion tensor  for the resolution of the field equations in a stationary and axisymmetric space-time is
twenty four, which significantly increases the mathematical complexity of the problem.

In the slow rotation approximation, the line element~\eqref{axi_line} is reduced to an expression that extends the metric tensor of the static and spherically symmetric solutions of cubic PG theory, by adding on top of the metric function~\eqref{formPsi} three arbitrary functions $\{f_{i}\}_{i=1}^{3}$ that depend on $r$ and $\vartheta$:
\begin{align}\label{eq:metric}
    ds^2=&\;\bigl(\Psi(r)+a\, f_1(r,\vartheta)\bigr)dt^2-\Bigl(\frac{1}{\Psi(r)}-a\, \frac{f_2(r,\vartheta)}{\Psi^{2}(r)}\Bigr)d r^2-r^2 d\vartheta^2- r^2 \sin^2\vartheta \,d\varphi^2\nonumber\\
    &+2a\bigl(1-\Psi(r)-f_3(r,\vartheta)\bigr)\sin ^2\vartheta\, dt  d\varphi\,,
\end{align}
where the parameter $a$ refers to the conventional Kerr angular momentum per unit mass, which contributes only at leading (linear) order. In this regard, it is important to stress the extrinsic character of this parameter, which does not operate as a source of torsion and accordingly must be distinguished from the spin charge $\kappa_{\rm s}$ describing the intrinsic angular momentum~\cite{Hehl:2012pi}.

On the other hand, the torsion tensor of the general static and spherically symmetric solution is extended with twenty four corrections depending on the radial and polar coordinates, which can be conveniently parametrised in the spin connection as follows\footnote{Note that in this parametrisation the respective functions $q_{i}(r,\vartheta)$ represent stationary and axisymmetric corrections extending the form of the spin connection of the general static and spherically symmetric solution.}:
\begin{align}
    w^{\hat{0}\hat{1}}{}_\mu=&\;a q_1(r,\vartheta )dt+a q_2(r,\vartheta ) dr+a q_3(r,\vartheta )d\vartheta+a q_4(r,\vartheta) d\varphi\,,\\
     w^{\hat{0}\hat{2}}{}_\mu=&\;a q_5(r,\vartheta )dt+a q_6(r,\vartheta )dr+\frac{1}{2}\bigl(2a q_7(r,\vartheta )-1\bigr) d\vartheta+\frac{1}{2}\bigl(2a q_8(r,\vartheta )- N_2 \kappa_{\rm s} \sin \vartheta \bigr) d\varphi\,,\\
       w^{\hat{0}\hat{3}}{}_\mu=&\;a q_9(r,\vartheta)dt+a q_{10}(r,\vartheta )dr+\frac{1}{2}\bigl(2aq_{11}(r,\vartheta )+N_2 \kappa_{\rm s}\bigr) d\vartheta+\frac{1}{2}\bigl(2aq_{12}(r,\vartheta )-\sin \vartheta\bigr)d\varphi\,,\\
          w^{\hat{1}\hat{2}}{}_\mu=&\;aq_{13}(r,\vartheta )dt+aq_{14}(r,\vartheta )dr+\frac{1}{2}\bigl(2a q_{15}(r,\vartheta )+1\bigr)d\vartheta+\frac{1}{2}\bigl(2a q_{16}(r,\vartheta )-N_2 \kappa_{\rm s}\sin \vartheta \bigr)d\varphi\,,\\
            w^{\hat{1}\hat{3}}{}_\mu=&\;a q_{17}(r,\vartheta )dt+aq_{18}(r,\vartheta )dr+\frac{1}{2}\bigl(2aq_{19}(r,\vartheta )+N_2 \kappa_{\rm s}\bigr) d\vartheta+\frac{1}{2}\bigl(2a q_{20}(r,\vartheta )+\sin \vartheta\bigr)d\varphi\,,\\
              w^{\hat{2}\hat{3}}{}_\mu=&\;\frac{1}{2r}\bigl[2arq_{21}(r,\vartheta )-\left(2 N_1+N_2\right) \kappa_{\rm s}\bigr]dt+\frac{1}{2r\Psi(r)}\bigl[2ar\Psi(r)q_{22}(r,\vartheta )+\left(2 N_1+N_2\right) \kappa_{\rm s}\bigr]dr+a q_{23}(r,\vartheta )d\vartheta\nonumber\\
              &+\bigl(a q_{24}(r,\vartheta )+\cos \vartheta\bigr)d\varphi\,,
              \end{align}
whereas the corresponding tetrad field reads:
\begin{eqnarray}
\vartheta^a{}_\mu =&\left(
\begin{array}{cccc}
\frac{1}{2} (\Psi(r)+1)+ \frac{a (\Psi(r)+1) f_{1}(r,\vartheta)}{4 \Psi(r)} & \frac{\Psi(r)-1}{2 \Psi(r)}-\frac{a (\Psi(r)-1) f_{2}(r,\vartheta)}{4 \Psi(r)^2} & 0 & -\frac{a (\Psi(r)+1) \sin ^2\vartheta}{2 \Psi(r)} (f_{3}(r,\vartheta)+\Psi(r)-1) \\
 \frac{1}{2} (\Psi(r)-1)+\frac{a (\Psi(r)-1) f_{1}(r,\vartheta)}{4 \Psi(r)} & \frac{\Psi(r)+1}{2 \Psi(r)}-\frac{a (\Psi(r)+1) f_{2}(r,\vartheta)}{4 \Psi(r)^2} & 0 & -\frac{a (\Psi(r)-1) \sin ^2\vartheta  }{2 \Psi(r)} (f_{3}(r,\vartheta)+\Psi(r)-1)\\
 0 & 0 & r & 0 \\
 0 & 0 & 0 & r \sin \vartheta  \\
\end{array}
\right)\,.
\end{eqnarray}

Therefore, the total number of dof in the metric and torsion tensors that will be considered for the resolution of the field equations in the slowly rotating space-time is twenty seven.

\subsection{Analytical solutions}\label{sec:solutions}

In this section, we derive analytical slowly rotating solutions for the degenerate model described by Expression~\eqref{cubicPGmodel_irreducible_torsion}, with the Lagrangian coefficients~\eqref{Lag_coeff1}-\eqref{Lag_coeff24} and $N_{2}=-\,N_{1}$. For simplicity, we can assume that the mass $m_{t}$ of the tensor mode vanishes, which can be directly obtained from Expression~\eqref{Lag_coeff24} by setting $h_{6}=0$ and $h_{25}=-\,d_{1}/3$.

By expanding the connection equation~\eqref{FieldEqSpin} to first order in the angular momentum $a$, one obtains $24$ second-order partial differential equations for the axisymmetric functions. These equations involve 24 dof from the torsion sector, which in the previous section we denoted by $q_i(r,\vartheta)$, with $i = 1, \dots, 24$; together with $3$ dof from the metric sector, denoted by $f_i(r,\vartheta)$, with $i = 1, \dots, 3$. We parametrise these functions by separating them into a part independent of the intrinsic spin charge $\kappa_{\rm s}$ and a part that depends linearly on this parameter:
\begin{align}
        q_i(r,\vartheta)&=\mathring{q}_i(r,\vartheta)+\kappa_{\rm s} \bar{q}_i(r,\vartheta)\,,\qquad i = 1,\dots, 24\,,\\
           f_i(r,\vartheta)&=\kappa_{\rm s} \bar{f}_i(r,\vartheta)\,,\qquad\, \,\,\,\,\,\,\,\,\,\,\,\,\,\,\,\,\,\,\,\,\,\,\,\, i = 1, \dots, 3\,,
    \end{align}
in such a way that the metric functions $f_i(r,\vartheta)$ vanish in the absence of $\kappa_{\rm s}$, where the dynamical behaviour of torsion is switched off and the Riemannian structure of the solution is unaffected. In analogy with the static and spherically symmetric solution, the $24$ dof of torsion are split into two sets: one comprising $24$ modes with vanishing intrinsic spin charge $\kappa_{\rm s}$ and another one comprising $24$ modes proportional to $\kappa_{\rm s}$. Then, the connection equation~\eqref{FieldEqSpin} takes the following form in the slow rotation approximation:
\begin{eqnarray}
  G_{0,i}(r,\vartheta) + G_{1,i}(r,\vartheta)\,\kappa_{\rm s} + G_{2,i}(r,\vartheta)\,\kappa_{\rm s}^2 + G_{3,i}(r,\vartheta)\,\kappa_{\rm s}^3 + G_{4,i}(r,\vartheta)\,\kappa_{\rm s}^4 = 0\,, \qquad i = 1, \dots, 24\,,\label{step0}
\end{eqnarray}
where the explicit form of the functions $G$ is omitted for brevity, since they consist of lengthy differential expressions for the variables $q_i(r,\vartheta)$ and $f_i(r,\vartheta)$. For simplicity, we will assume that each contribution is zero independently, i.e.
    \begin{eqnarray}
        G_{0,i}(r,\vartheta)=G_{1,i}(r,\vartheta)=G_{2,i}(r,\vartheta)=G_{3,i}(r,\vartheta)=G_{4,i}(r,\vartheta)=0\,, \qquad i = 1, \dots, 24\,.
    \end{eqnarray}
Then, there is a set of constraints arising from the last three conditions $G_{2,i}(r,\vartheta)=G_{3,i}(r,\vartheta)=G_{4,i}(r,\vartheta)=0$ that directly restrict the metric functions as:
\begin{eqnarray}
        \bar{f}_2(r,\vartheta) = -\,\bar{f}_1(r,\vartheta) = \frac{\Psi(r)\,\big(\bar{f}_{2,B}(\vartheta) + K_1 r^3\big)}{r}\,, 
        \qquad \bar{f}_3(r,\vartheta) = 0\,, \label{step3}
\end{eqnarray}
where $K_1$ is an integration constant, $\bar{f}_{2,B}(\vartheta )$ is an arbitrary function of $\vartheta$ and $\Psi(r) = 1 - 2m/r$. On the other hand, the rest of constraints provided by the conditions $G_{3,i}(r,\vartheta) = G_{4,i}(r,\vartheta) = 0$ set the following functions of the spin connection as:
\begin{align}
\bar{q}_{3}(r,\vartheta) &= -\,\frac{r}{2}\,(\bar{q}_{13}-\bar{q}_{5}+\bar{q}_{6})-\frac{r}{2\Psi}\,\bar{q}_{5}\,, \\
\bar{q}_{4}(r,\vartheta) &= -\,\frac{r}{2}\,(\bar{q}_{10}+\bar{q}_{17}-\bar{q}_{9})\sin\vartheta - \frac{r}{2\Psi}\sin\vartheta\,\bar{q}_{9}\,, \\
\bar{q}_{14}(r,\vartheta) &= \bar{q}_{6} + \frac{1}{\Psi}\left(\bar{q}_{5}-\bar{q}_{13}\right)\,, \\
\bar{q}_{15}(r,\vartheta) &= \bar{q}_{7}-r\bar{q}_{2} + \frac{\bar{f}_2+r \partial_r \bar{f}_2-2 r \bar{q}_1}{2 \Psi }-\frac{r \Psi ' \bar{f}_2}{2 \Psi^2}\,, \\
\bar{q}_{18}(r,\vartheta) &= \bar{q}_{10} + \frac{1}{\Psi}\left(\bar{q}_{9}-\bar{q}_{17}\right), \\
\bar{q}_{19}(r,\vartheta) &= \bar{q}_{11} + \csc \vartheta\,(\bar{q}_{8}-\bar{q}_{16})\,, \\
\bar{q}_{20}(r,\vartheta) &= \bar{q}_{12}-r\sin\vartheta\,\bar{q}_{2} + \frac{\sin \vartheta \left(\bar{f}_2+r\partial_r\bar{f}_2-2 r\bar{q}_1\right) }{2 \Psi }-\frac{r \sin \vartheta  \Psi '\bar{f}_2}{2 \Psi^2} \,, \\
\bar{q}_{22}(r,\vartheta) &= \frac{1}{r}\csc \vartheta\,(\bar{q}_{16}-\bar{q}_{8}) - \frac{1}{\Psi}\,\bar{q}_{21}\,.
\end{align}
For simplicity in the notation, we omit the explicit dependence on $(r,\vartheta)$ in the expressions.
Likewise, the rest of constraints given by the condition $G_{2,i}(r,\vartheta)=0$ restrict the form of the following functions:
\begin{align}
\mathring{q}_{10}(r,\vartheta) =&\; 
\frac{1}{2N_{1}r\Psi^2(1+\Psi)}\Big\{\Psi ^2 \left[7 r \partial_r\partial_\vartheta\bar{f}_2(r,\vartheta )+4 \partial_\vartheta\bar{f}_2+3 r^2 \Psi ' \left(\bar{q}_6-\bar{q}_5+\bar{q}_{13}\right)-3 r \bar{q}_5-4 N_1 \csc \vartheta  \mathring{q}_4\right]
    \nonumber \\
& +\Psi  \left[\Psi ' \left(r \left(3 r \bar{q}_5-7 \partial_\vartheta\bar{f}_2\right)+N_1 \sin \vartheta \right)+\partial_\vartheta\bar{f}_2+r \partial_r\partial_\vartheta\bar{f}_2-2 N_1 r \mathring{q}_9\right]+r \Psi ^3 \left(3 \bar{q}_5-3 \bar{q}_6-3 \bar{q}_{13}+2 N_1 \mathring{q}_{18}\right)\nonumber\\
&-\Psi ' \left(N_1 \sin \vartheta+r \partial_\vartheta\bar{f}_2\right)\Big\}\,, \\
\mathring{q}_{14} =&\; 
\mathring{q}_{6} 
+ \frac{1}{\Psi}\left(\mathring{q}_{5}-\mathring{q}_{13}\right)\,, \\
\mathring{q}_{16}(r,\vartheta) =&\; 
\sin\vartheta\left(\mathring{q}_{11}-\mathring{q}_{19}\right)
+ \mathring{q}_{8}\,, \\
\mathring{q}_{17}(r,\vartheta) =&\; 
\frac{1}{2N_{1}r \Psi^2(1+\Psi)}\Big\{\Psi ^3 \left[5 r \partial_r\partial_\vartheta\bar{f}_2+2 \partial_\vartheta\bar{f}_2+3 r^2 \Psi ' \left(\bar{q}_6-\bar{q}_5+\bar{q}_{13}\right)-3 r \bar{q}_5-2 N_1 \left(r \mathring{q}_{18}-r \mathring{q}_9+2 \csc \vartheta  \mathring{q}_4\right)\right]\nonumber\\
&+\Psi ^2 \left[\Psi ' \left(r \left(3 r \bar{q}_5-5 \partial_\vartheta\bar{f}_2\right)+3 N_1 \sin \vartheta \right)-\partial_\vartheta\bar{f}_2-r \partial_r\partial_\vartheta\bar{f}_2 \right]+\Psi  \Psi ' \left(r \partial_\vartheta\bar{f}_2+N_1 \sin \vartheta \right)\nonumber\\
&-3 r \Psi ^4 \left(\bar{q}_6-\bar{q}_5+\bar{q}_{13}\right) \Big\}\,, \\
\mathring{q}_{20}(r,\vartheta) =&\; 
\frac{2\Psi-1}{N_{1}}\big(\bar{q}_{8}-\bar{q}_{16}\big)
- r\sin\vartheta\,\mathring{q}_{2}
+ \mathring{q}_{12}
- \frac{r\sin\vartheta}{\Psi}\,\mathring{q}_{1}+\frac{\left(r \Psi '+\Psi -1\right) \left(\bar{q}_{16}-\bar{q}_8\right)}{N_1}\,, \\
\mathring{q}_{21}(r,\vartheta) =&\; 
\frac{1}{2N_{1}r^{2}}\Big\{\Psi \left[4 r^2 \partial_r\bar{f}_2-6 r^3 \partial_r\partial_r\bar{f}_2+4 r \bar{f}_2+2 N_1 \left(r \mathring{q}_{11}-r^2 \mathring{q}_{22}-r \mathring{q}_{19}+2 \cos \vartheta \right)\right]+6 r^3 \Psi '' \bar{f}_2-\frac{11 r^3 \Psi '^2 \bar{f}_2}{\Psi }\nonumber\\
&+r^2 \Psi '\left(11 r \partial_r\bar{f}_2-2 \bar{f}_2\right)-4 N_1 \cos \vartheta  \Big\}\,, \\
\mathring{q}_{3}(r,\vartheta) =& 
-\frac{r}{4N_{1}\Psi}\big\{3 r \Psi ' \bar{q}_9+2 N_1 \mathring{q}_5+\Psi\left[3\left(r \Psi'-\Psi\right)\left(\bar{q}_{10}-\bar{q}_9+\bar{q}_{17}\right)-3 \bar{q}_9+2 N_1 \left(\mathring{q}_6-\mathring{q}_5+\mathring{q}_{13}\right)\right]\big\}\,, \\
\mathring{q}_{7}(r,\vartheta) =&\; 
\mathring{q}_{15}
+ r\mathring{q}_{2}
+ \frac{r}{\Psi}\,\mathring{q}_{1}
- \frac{1}{N_{1}}\csc\vartheta\,(r \Psi'-\Psi)\big(\bar{q}_{16}-\bar{q}_{8}\big)\,.
\end{align}

Once the conditions $G_{2,i}(r,\vartheta) = G_{3,i}(r,\vartheta) = G_{4,i}(r,\vartheta) = 0$ of the connection field equation are completely solved, it is possible to follow the same procedure with the tetrad field equation~\eqref{fieldeqTetrad}, which straightforwardly vanishes the constant $K_{1}$ and the arbitrary function $\bar{f}_{2,B}(\vartheta)$: \begin{equation}
    K_{1} = \bar{f}_{2,B}(\vartheta) = 0\,,
\end{equation}
i.e. vanishing the metric functions $\bar{f}_{1}(r,\vartheta)=\bar{f}_{2}(r,\vartheta)=0$ and fixing the geometry to the slowly rotating Kerr space-time.

At this point, two sets of conditions remain to solve the connection field equation completely: $G_{0,i}(r,\vartheta)=0$ and $G_{1,i}(r,\vartheta)=0$. Unfortunately, there is no a general way to solve these conditions explicitly, but we shall use an angular ansatz for the remaining $32$ functions that have not been fixed yet by the field equations, which turns out to provide nontrivial solutions.

First, for the $8$ functions associated with the component $\omega_{\varphi}$ of the spin connection, we assume:
\begin{align}
        \mathring{q}_i(r,\vartheta) =&\; \mathring{q}_{i,R}(r) + \mathring{q}_{i,A}(r)\sin\vartheta + \mathring{q}_{i,B}(r)\cos\vartheta + \mathring{q}_{i,C}(r)\sin\vartheta\cos\vartheta + \mathring{q}_{i,D}(r)\sin^2\vartheta+ \mathring{q}_{i,E}(r)\cos^2\vartheta \,,\\
        \bar{q}_j(r,\vartheta) =&\;  \bar{q}_{j,R}(r) + \bar{q}_{j,A}(r)\sin\vartheta + \bar{q}_{j,B}(r)\cos\vartheta + \bar{q}_{j,C}(r)\sin\vartheta\cos\vartheta + \bar{q}_{j,D}(r)\sin^2\vartheta+ \bar{q}_{j,E}(r)\cos^2\vartheta\,,
        \end{align}
with $i=4,8,12,24$ and $j=8,12,16,24$. Then, for the $17$ functions associated with the components $\omega_{t}$ and $\omega_{r}$:
\begin{align}
    \mathring{q}_{k}(r,\vartheta) =&\;  \mathring{q}_{k,R}(r) + \mathring{q}_{k,A}(r)\sin\vartheta + \mathring{q}_{k,B}(r)\cos\vartheta \,,\quad k=1,2,5,6,9,13,18,22\,,\\
     \bar{q}_{l}(r,\vartheta) =& \; \bar{q}_{l,R}(r) + \bar{q}_{l,A}(r)\sin\vartheta + \bar{q}_{l,B}(r)\cos\vartheta  \,,\quad l=1,2,5,6,9,10,13,17,21\,,
\end{align}
whereas for the last $7$ functions that are associated with the component $\omega_{\vartheta}$:
\begin{align}
    \mathring{q}_{n}(r,\vartheta) =&\;  \mathring{q}_{n,R}(r) + \mathring{q}_{n,A}(r)\sin\vartheta + \mathring{q}_{n,B}(r)\cos\vartheta \,,\quad n=11,15,19,23\,,\\
     \bar{q}_{p}(r,\vartheta) =&\;  \bar{q}_{p,R}(r) + \bar{q}_{p,A}(r)\sin\vartheta + \bar{q}_{p,B}(r)\cos\vartheta  \,,\quad p=7,11,23\,.
\end{align}
Thereby, our ansatz introduces a total of $130$ arbitrary functions of the radial coordinate. By substituting the ansatz into the field equations, the angular dependence is now explicitly reduced to the trigonometric functions above and the remaining unknowns depend solely on the radial coordinate $r$. Hence, in order for the equations to hold for all $\vartheta$, the coefficients in front of each trigonometric function must vanish separately. This leads to a large system of differential equations for the radial functions introduced earlier. While solving the full system is in general tedious and cumbersome, almost all of the equations can be solved systematically, with only three remaining differential equations that are decoupled. Hence, the system is solvable, although the general solution is rather involved. Likewise, it is also important to note that, once the connection equation is completely solved, the tetrad equation is automatically satisfied for $\Psi(r) = 1-2m/r$.

Although the final form of the solution is quite involved, for simplicity we set all of the integration constants and arbitrary functions that appear in the solution by this procedure to zero, except for the ones providing contributions of order \(a \kappa_{\rm s}\) in the gravitational action~\eqref{cubicPGmodel_irreducible_torsion}. With this simplification, the solution of the field equations takes the form:
\begin{align}
    f_1(r,\vartheta)&=f_2(r,\vartheta)=f_3(r,\vartheta)=0\,, \quad q_1(r,\vartheta)=-\,\Psi (r)q_2(r,\vartheta)=\biggl[\frac{ r\Psi'(r)+\Psi(r)\big(1+\Psi(r)\big)}{4 r^2 \Psi^{2} (r)}\biggr]N_1\kappa_{\rm s} \cos \vartheta \,,\\
    q_3(r,\vartheta)&=rq_5(r,\vartheta)=-\,r q_{14}(r,\vartheta)=\frac{N_1 \kappa_{\rm s} \sin \vartheta}{2 r \Psi (r)}\,, \quad q_4(r,\vartheta)=\frac{\sin ^2\vartheta }{2 r}\,,\quad q_6(r,\vartheta)=-\,\Big(\frac{1+\Psi (r)}{2 r^2 \Psi^{2} (r)}\Big)N_1\kappa_{\rm s} \sin \vartheta \,,\\
    q_7(r,\vartheta)&=q_{15}(r,\vartheta)=\frac{N_1 \big(1+\Psi (r)\big) \kappa_{\rm s} \cos \vartheta }{4 r \Psi (r)}\,, \quad q_8(r,\vartheta)=-\,q_{11}(r,\vartheta)\sin\vartheta=-\,F(r,\vartheta)\sin \vartheta \,,\quad q_9(r,\vartheta)=0\,,\\
    q_{10}(r,\vartheta)&=\frac{\big(\Psi^{2} (r)-r\Psi'(r)\big) }{2r^2 \Psi^{2} (r)}\sin \vartheta\,, \quad q_{12}(r,\vartheta)=q_{20}(r,\vartheta)=\frac{N_1 \big(1+\Psi(r)\big) \kappa_{\rm s} }{4 r \Psi (r)}\sin \vartheta  \cos \vartheta \,,\quad q_{13}(r,\vartheta)=0\,,\quad \\
    q_{16}(r,\vartheta)&=-\,q_{19}(r,\vartheta)\sin\vartheta =\Big[\frac{\cos\vartheta}{r}\Big(\frac{1}{\Psi(r)}-1\Big)-F(r,\vartheta)\Big]\sin\vartheta\,,\quad q_{17}(r,\vartheta)=0\,, \quad q_{23}(r,\vartheta)=\frac{\sin \vartheta }{2 r} \\
    q_{18}(r,\vartheta)&=\frac{\Psi^{2}(r)+r\Psi'(r)}{2 r^2 \Psi^{2} (r)}\sin \vartheta \,, \quad q_{22}(r,\vartheta)=\frac{\big(2\Psi^{2}(r)+1\big)\cos\vartheta-2\Psi (r)\big(r F(r,\theta )+2\cos\vartheta\big)}{2 r^2 \Psi^{2} (r)}\\
    q_{21}(r,\vartheta)&=\frac{2 r\Psi(r) F(r,\theta )+\big(2\Psi(r)-1\big) \cos \vartheta }{2 r^2\Psi(r)}\,, \quad q_{24}(r,\vartheta)=\frac{\big(\Psi (r)-1\big)}{2r\Psi (r)}N_1\kappa_{\rm s} \sin ^2\vartheta \,,
\end{align}
where $
    F(r,\vartheta)=\frac{1}{r}[\sin \vartheta  \mathring{q}_{19,A}(r)+\cos \vartheta  \mathring{q}_{11,B}(r)-\Psi (r) \mathring{q}_{22,R}(r)]$. Note that under the angular ansatz the $\vartheta$-dependence of this particular function is fixed a priori, so the arbitrariness in $F(r,\vartheta)$ cannot be seen at the ansatz level. However, after obtaining the explicit solution, one can verify that the field equations remain satisfied if this quantity is promoted to an arbitrary function $F(r,\vartheta)$, while the rest of terms of the solution remain unchanged. Then, by considering the solution with this arbitrary function, it is straightforward to check the Lagrangian density of the solution displays a gravitational SOI of the form:
\begin{equation}
    \mathcal{L}=\frac{d_1 N_1^2 \kappa_{\rm s}^2}{8 \pi  r^4}+\mathcal{L}_{\rm SOI}\,,
\end{equation}
where
\begin{equation}
    \mathcal{L}_{\rm SOI}=\frac{a d_1 N_1 \kappa_{\rm s}} {2 \pi  r^4}\Big(r \partial_{r}F(r,\vartheta)-F(r,\vartheta )+\frac{   r \Psi '(r)+2\Psi (r)-4\Psi^{2}(r)}{2  r \Psi^{2} (r)}\, \cos \vartheta\Big)\,.
\end{equation}

Re-expressing the arbitrary function as
\begin{equation}
    F(r,\vartheta) = \frac{\bigl(1-2\Psi(r)\bigr)}{2r\Psi(r)}\cos\vartheta+r\int r^{2}G(r,\vartheta)\,dr\,,
\end{equation} the Lagrangian density of the interaction reads:
\begin{equation}
    \mathcal{L}_{\rm SOI}=\frac{ d_{1}N_{1}a\kappa_{\rm s}G(r,\vartheta)}{2\pi}\,.
\end{equation}
Thereby, the function $G(r,\vartheta)$ directly provides the interaction between the intrinsic and extrinsic angular momentum parameters in the Lagrangian density, which remains general for the solution. Thus, for instance, setting the form of this function as
\begin{equation}
    G(r,\vartheta) = \frac{1}{\sqrt{-\,g}}\frac{\Psi'(r)}{r}\cos\vartheta\,,
\end{equation}
would provide a term analogous to the Thomas precession of atomic systems, but in this case with a purely gravitational origin. In particular, it turns out this function is encoded solely in the axial mode of the torsion tensor of the solution, so the trajectories of Dirac particles minimally coupled to torsion will accordingly experience deviations from the geodesic motion, which can already be measured in the semiclassical limit by an acceleration of the form~\cite{Audretsch:1981xn,Cembranos:2018ipn}:
\begin{equation}
    u^{\lambda}\nabla_{\lambda}u_{\mu} = \frac{1}{4m_{\rm s}}\hat{R}_{\lambda\rho\mu\nu}\bar{b}_{0}\sigma^{\lambda\rho}b_{0}u^{\nu}\,,
\end{equation}
where $u_{\mu}$ represents the four-velocity of the particle, $b_{0}$ its normalised state, $m_{\rm s}$ its mass, $\sigma^{\lambda\rho}$ the spin matrices and $\hat{R}_{\lambda\rho\mu\nu}$ the part of the RC curvature tensor that includes corrections from the axial mode alone. In any case, it is important to stress the interaction cannot prevent the singular behaviour of those Dirac particles crossing the event horizon of the solution, in virtue of the strongly asymptotically predictable nature of the slowly rotating Kerr black hole~\cite{Cembranos:2016xqx,Cembranos:2019mcb}.

\section{Conclusions}\label{sec:conclusions}

In this letter, we have investigated the gravitational SOI in the framework of PG theory. It is well-known that this theory successfully describes the gravitational effects of the intrinsic and extrinsic parts of the canonical angular momentum tensor~\cite{Hehl:2012pi}, but the implications of the interaction between these two quantities at macroscopical scales have not been explored in the literature so far. In particular, it is important to stress that the metric structure of a stationary and axisymmetric space-time is not fixed in general by the field equations to the conventional Kerr solution. This means that the corresponding extensions of the static and spherically symmetric configurations endowed with a spin charge to the stationary and axisymmetric space-time do not satisfy in general the Kerr hypothesis~\cite{Herdeiro:2022yle,Shahzadi:2023act,Yang:2025aro}, but the corresponding interaction with the extrinsic angular momentum can potentially lead to a new geometry.

In any case, the highly nonlinear character of the field equations of PG theory and the large number of dof present in the stationary and axisymmetric space-time constitute a major challenge for this task. Thereby, in this work we have initially focused on the search of black hole solutions that at least can display such an interaction in the gravitational action, regardless it does not provide any modification in the metric tensor. Thus, we have considered a degenerate model of cubic PG theory, which provides static and spherically symmetric solutions with a spin charge that does not affect the Schwarzschild geometry, and we have analytically solved its field equations in the slow rotation approximation. As expected, the corresponding slowly rotating solutions also respect the Kerr metric structure, but they explicitly display the SOI in the gravitational action, which is provided by a general function depending on the radial and polar coordinates.

From this result, it would be interesting to perform further analyses to clarify if certain values of this interaction can be particularly favored from the phenomenological point of view. Likewise, the search of a solution within a nondegenerate model of PG theory is especially relevant. In particular, the nondegenerate models of cubic PG theory provide static and spherically symmetric solutions beyond the Schwarzschild space-time~\cite{Bahamonde:2024sqo}, so that the corresponding stationary and axisymmetric extensions can potentially give rise to a new geometry beyond the Kerr black hole, where the gravitational SOI could play a significant role in the geometrical structure of the solution (e.g. enabling a regular black hole solution beyond the slow rotation approximation). The main obstruction for this task is the computational intractability of the field equations of the theory, which indicates that the application of different constraints and/or hidden symmetries is expected to play a significant role to simplify and solve this problem. In this regard, we want to emphasise that the torsion field associated with the intrinsic angular momentum does not constitute any tensor field but the translational differential form derived from a particular gauge connection, such as the one introduced in PG theory (see also~\cite{Tseytlin:1981nu} and~\cite{Koivisto:2025ryb} for alternative formulations based on the $S_{10}$ and $Spin(4)$ groups). Accordingly, the field equations that must be considered to describe any physical implication of torsion (including a gravitational SOI like the one investigated in our work) must respect the corresponding translational gauge invariance, which means that only a solution of these field equations can truly represent a physical configuration with a gravitational SOI related to torsion (see Fallacy 5 and Fallacy 6 in~\cite{Blagojevic:2013xpa}).

\noindent
\section*{Acknowledgements}

S.B. is supported by “Agencia Nacional de Investigación y Desarrollo” (ANID), Grant “Becas Chile postdoctorado al extranjero” No. 74220006 and also by Institute for Basic Science (IBS-R018-D3). The work of J.G.V. is supported by the Institute for Basic Science (IBS-R003-D1).

\newpage

\appendix

\section{Field equations of PG theory with cubic order invariants defined from the torsion and curvature tensors}\label{appe1}

Variation of the action~\eqref{cubicPGmodel_fulltorsion} imposes the tensors $E^{\lambda\mu\nu}$ and $E^{\mu\nu}$ to vanish, according to the principle of least action. Specifically, the tensor $E^{\lambda\mu\nu}$ provides the connection field equation and acquires the form:
\begin{equation}
    E^{\lambda\mu\nu}=\frac{1}{2}\bigl(X^{\lambda \nu \mu }- X^{\lambda \mu \nu }  + X^{\mu \lambda \nu } -  X^{\mu \nu \lambda }\bigr)+2\bigl(K^{\mu }{}_{\alpha \kappa } Y^{\lambda \alpha \nu \kappa } - K^{\lambda }{}_{\alpha \kappa } Y^{\mu \alpha \nu \kappa } - \nabla_{\alpha }Y^{\mu \lambda \nu \alpha }\bigr)=0\,,\label{FieldEqSpin}
\end{equation}
where we have defined
\begin{align}
X^{\lambda\mu\nu}&=U^{\lambda[\mu\nu]}\,,\\
    Y^{\lambda\rho\mu\nu}&=W^{[\lambda\rho][\mu\nu]}\,,
\end{align}
being at the same time the tensor $U_{\lambda \rho \mu }$ expressed as
\begin{align}
    U_{\lambda \rho \mu } =&\;\frac{1}{2} m_{T}^2 \big(g_{\lambda \rho } T_{\mu } -  g_{\lambda \mu } T_{\rho }\big)-2 m_{S}^2 \big(T_{\lambda \mu \rho } +  T_{\mu \rho\lambda} + T_{\rho \lambda \mu }\big) + \frac{1}{3} m_{t}^2 \big(2 T_{\lambda \mu \rho } + T_{\mu \lambda \rho } +  T_{\rho\mu \lambda} + g_{\lambda \mu } T_{\rho }-  g_{\lambda \rho } T_{\mu }\big)\nonumber\\
    &+\bar{h}_{1}^{} \big(\tilde{R}^{\alpha \kappa }{}_{\mu \rho } + \tilde{R}_{\mu \rho }{}^{\alpha \kappa }\big) T_{\lambda \alpha \kappa } + 
2 \bar{h}_{2}^{} \tilde{R}_{\rho }{}^{\kappa }{}_{\mu }{}^{\alpha } T_{\lambda \alpha \kappa } + 
\bar{h}_{3}^{} \big(\tilde{R}^{\alpha \kappa }{}_{\mu \rho } T_{\alpha \lambda \kappa } + \tilde{R}_{\lambda \rho }{}^{\alpha \kappa } T_{\mu \alpha \kappa }\big)\nonumber\\
& - 
\bar{h}_{4}^{} \big(\tilde{R}_{\rho }{}^{\alpha }{}_{\mu }{}^{\kappa } T_{\alpha \lambda \kappa } +  \tilde{R}_{\lambda }{}^{\kappa }{}_{\rho }{}^{\alpha } T_{\mu \alpha \kappa }\big)- 
\bar{h}_{5}^{} \big( \tilde{R}_{\rho }{}^{\kappa }{}_{\mu }{}^{\alpha } T_{\alpha \lambda \kappa } +  \tilde{R}_{\rho }{}^{\kappa }{}_{\lambda }{}^{\alpha } T_{\mu \alpha \kappa }\big) + 
\bar{h}_{6}^{} \big(\tilde{R}_{\mu \rho }{}^{\alpha \kappa } T_{\alpha \lambda \kappa } + \tilde{R}^{\alpha \kappa }{}_{\lambda \rho } T_{\mu \alpha \kappa }\big) \nonumber\\
&+ 
\bar{h}_{7}^{} \big(g_{\lambda \mu } \tilde{R}_{\rho }{}^{\alpha \kappa \theta } T_{\alpha \theta \kappa } -  \tilde{R}_{\lambda }{}^{\alpha }{}_{\mu \rho } T_{\alpha }\big) + 
\bar{h}_{8}^{} \big( \tilde{R}_{\rho }{}^{\alpha }{}_{\lambda \mu } T_{\alpha }- g_{\lambda \mu } \tilde{R}_{\rho }{}^{\theta \alpha \kappa } T_{\alpha \theta \kappa }\big) + 
\bar{h}_{9}^{} \big( \tilde{R}_{\lambda \rho \mu }{}^{\alpha } T_{\alpha }- g_{\lambda \mu } \tilde{R}^{\alpha \theta }{}_{\rho }{}^{\kappa } T_{\alpha \theta \kappa }\big)\nonumber\\
& + 
\bar{h}_{10}^{} \big(g_{\lambda \mu } \tilde{R}^{\kappa \theta }{}_{\rho }{}^{\alpha } T_{\alpha \theta \kappa } -  \tilde{R}_{\mu \rho \lambda }{}^{\alpha } T_{\alpha }\big) + 
\bar{h}_{11}^{} \big(\tilde{R}^{\alpha \kappa }{}_{\lambda \rho } + \tilde{R}_{\lambda \rho }{}^{\alpha \kappa }\big) T_{\alpha \mu \kappa } + 
2 \bar{h}_{12}^{} \tilde{R}_{\lambda }{}^{\alpha }{}_{\rho }{}^{\kappa } T_{\alpha \mu \kappa } \nonumber\\
&+ 
\bar{h}_{13}^{} \big(\tilde{R}_{\lambda }{}^{\kappa }{}_{\rho }{}^{\alpha } + \tilde{R}_{\rho }{}^{\alpha }{}_{\lambda }{}^{\kappa }\big) T_{\alpha \mu \kappa } + 
2 \bar{h}_{14}^{} \tilde{R}_{\rho }{}^{\kappa }{}_{\lambda }{}^{\alpha } T_{\alpha \mu \kappa }+ 
\bar{h}_{15}^{} \big(\tilde{R}_{\alpha \lambda } + \tilde{R}_{\lambda \alpha }\big) T^{\alpha }{}_{\mu \rho } + 
\bar{h}_{16}^{} \big( \tilde{R}_{\alpha \lambda } T_{\rho \mu }{}^{\alpha }- \tilde{R}_{\rho \alpha } T^{\alpha }{}_{\lambda \mu }\big)  \nonumber\\
&+ 
\bar{h}_{17}^{} \big( \tilde{R}_{\lambda \alpha } T_{\rho \mu }{}^{\alpha }- \tilde{R}_{\alpha \rho } T^{\alpha }{}_{\lambda \mu }\big) + 
\bar{h}_{18}^{} \big(\tilde{R}_{\alpha \rho } + \tilde{R}_{\rho \alpha }\big) T_{\lambda \mu }{}^{\alpha } + 
\bar{h}_{19}^{} \big(\tilde{R}_{\alpha \rho } + \tilde{R}_{\rho \alpha }\big) T_{\mu \lambda }{}^{\alpha } 
- \bar{h}_{20}^{} g_{\lambda \mu } \big(\tilde{R}_{\alpha \rho } + \tilde{R}_{\rho \alpha }\big) T^{\alpha }\nonumber\\
& + 
\bar{h}_{21}^{} \big(g_{\lambda \mu } \tilde{R}_{\alpha \kappa } T^{\alpha }{}_{\rho }{}^{\kappa } -  \tilde{R}_{\lambda \rho } T_{\mu }\big) + 
\bar{h}_{22}^{} \big(g_{\lambda \mu } \tilde{R}_{\alpha \kappa } T^{\kappa }{}_{\rho }{}^{\alpha } -  \tilde{R}_{\rho \lambda } T_{\mu }\big) - 
\bar{h}_{23}^{} \big(g_{\lambda \mu } \tilde{R}_{\alpha \kappa } T_{\rho }{}^{\alpha \kappa } +  \tilde{R}_{\rho \mu } T_{\lambda }\big)\nonumber\\
& + 
2 \bar{h}_{24}^{} \tilde{R} T_{\lambda \mu \rho } + 
\bar{h}_{25}^{} \tilde{R} \big(T_{\mu \lambda \rho } -  T_{\rho \lambda \mu }\big) + 
\bar{h}_{26}^{} \tilde{R} \big(g_{\lambda \rho } T_{\mu } -  g_{\lambda \mu } T_{\rho }\big)  \,,
\end{align}
and the tensor $W_{\lambda \rho \mu \nu }$ as
\begin{align}
     W_{\lambda \rho \mu \nu } =&\;2 c_{1}^{} \tilde{R}_{\rho \lambda \nu \mu } - 2 c_{2}^{} \tilde{R}_{\nu \rho \lambda \mu }- (2 c_{1}^{} + c_{2}^{}) \tilde{R}_{\nu \mu \rho \lambda } + d_{1}^{} \big[g_{\nu \rho } \big(\tilde{R}_{\lambda \mu } -  \tilde{R}_{\mu \lambda }\big) + g_{\lambda \nu } \big(\tilde{R}_{\mu \rho } -  \tilde{R}_{\rho \mu }\big)\big]\nonumber\\
    &+\bar{h}_{1}^{} T_{\alpha \lambda \rho } T^{\alpha }{}_{\mu \nu } 
- \bar{h}_{2}^{} T_{\alpha \lambda \mu } T^{\alpha }{}_{\nu \rho }  
- \bar{h}_{3}^{} T_{\alpha \mu \nu } T_{\lambda \rho }{}^{\alpha } + 
\bar{h}_{4}^{} T_{\alpha \nu \rho } T_{\lambda \mu }{}^{\alpha } + 
\bar{h}_{5}^{} T_{\alpha \nu \rho } T_{\mu \lambda }{}^{\alpha } 
- \bar{h}_{6}^{} T_{\alpha \lambda \rho } T_{\mu \nu }{}^{\alpha }\nonumber\\
    & + 
\bar{h}_{7}^{} T_{\rho \mu \nu } T_{\lambda } 
- \bar{h}_{8}^{} T_{\mu \nu \rho } T_{\lambda } 
- \bar{h}_{9}^{} T_{\lambda \nu \rho } T_{\mu } + 
\bar{h}_{10}^{} T_{\nu \lambda \rho } T_{\mu } + 
\bar{h}_{11}^{} T_{\lambda \rho }{}^{\alpha } T_{\mu \nu \alpha } + 
\bar{h}_{12}^{} T_{\lambda \mu }{}^{\alpha } T_{\rho \nu \alpha } + 
\bar{h}_{13}^{} T_{\mu \lambda }{}^{\alpha } T_{\rho \nu \alpha } \nonumber\\
    &+ 
\bar{h}_{14}^{} T_{\mu \lambda }{}^{\alpha } T_{\nu \rho \alpha } + 
\bar{h}_{15}^{} g_{\nu \rho } T_{\lambda }{}^{\alpha }{}_{\kappa } T_{\mu \alpha }{}^{\kappa } + 
\bar{h}_{16}^{} g_{\nu \rho } T_{\alpha \lambda }{}^{\kappa } T_{\mu }{}^{\alpha }{}_{\kappa } + 
\bar{h}_{17}^{} g_{\nu \rho } T_{\alpha \mu }{}^{\kappa } T_{\lambda }{}^{\alpha }{}_{\kappa } + 
\bar{h}_{18}^{} g_{\nu \rho } T_{\alpha \lambda }{}^{\kappa } T^{\alpha }{}_{\mu \kappa }\nonumber\\
    & + 
\bar{h}_{19}^{} g_{\nu \rho } T_{\alpha \lambda }{}^{\kappa } T_{\kappa \mu }{}^{\alpha } + 
\bar{h}_{20}^{} g_{\nu \rho } T_{\lambda } T_{\mu } + 
\bar{h}_{21}^{} g_{\nu \rho } T_{\lambda \mu }{}^{\alpha } T_{\alpha } + 
\bar{h}_{22}^{} g_{\nu \rho } T_{\mu \lambda }{}^{\alpha } T_{\alpha } + 
\bar{h}_{23}^{} g_{\nu \rho } T_{\alpha \lambda \mu } T^{\alpha } \nonumber\\
    &+ 
\bar{h}_{24}^{} g_{\lambda \mu } g_{\nu \rho } T_{\alpha }{}^{\kappa \theta } T^{\alpha }{}_{\kappa \theta } + 
\bar{h}_{25}^{} g_{\lambda \mu } g_{\nu \rho } T_{\alpha }{}^{\kappa \theta } T_{\kappa }{}^{\alpha }{}_{\theta } + 
\bar{h}_{26}^{} g_{\lambda \mu } g_{\nu \rho } T_{\alpha } T^{\alpha }\,.
\end{align}

On the other hand, the tensor $E^{\mu\nu}$ provides the tetrad field equation and can be written as
\begin{equation}
    E^{\mu\nu}=2G^{\mu\nu}+\tilde{\mathcal{L}} \,g^{\mu \nu } + V^{\mu \nu } + 2 K^{\mu }{}_{\alpha \kappa } X^{\alpha \nu \kappa } + 2 \nabla_{\alpha}X^{\mu \nu \alpha }=0\,,\label{fieldeqTetrad}
\end{equation}
where $\tilde{\mathcal{L}}$ represents the Lagrangian density given by the quadratic and cubic order invariants in Expression~\eqref{cubicPGmodel_fulltorsion} and the tensor $V_{\mu\nu}$ reads
\begin{align}
V_{\mu \nu}=&\; d_{1}^{} \big(2 \tilde{R}_{\nu }{}^{\kappa }{}_{\mu }{}^{\alpha } \tilde{R}_{\alpha \kappa } -2 \tilde{R}_{\nu }{}^{\alpha }{}_{\mu }{}^{\kappa } \tilde{R}_{\alpha \kappa } - 2 \tilde{R}_{\alpha \mu } \tilde{R}^{\alpha }{}_{\nu } + 2 \tilde{R}^{\alpha }{}_{\mu } \tilde{R}_{\nu \alpha }\big)+4 c_{1}^{} \tilde{R}_{\theta \kappa \mu \alpha } \big(\tilde{R}^{\kappa \theta }{}_{\nu }{}^{\alpha } -  \tilde{R}_{\nu }{}^{\alpha \kappa \theta }\big)  \nonumber\\
&+ 2 c_{2}^{} \tilde{R}_{\theta \kappa \mu \alpha } \big(\tilde{R}^{\alpha \theta }{}_{\nu }{}^{\kappa } -  \tilde{R}_{\nu }{}^{\alpha \kappa \theta } + \tilde{R}_{\nu }{}^{\theta \kappa \alpha }\big)  + 4 m_{S}^2 \big(T_{\alpha \nu }{}^{\kappa } T^{\alpha }{}_{\mu \kappa } -  T^{\alpha }{}_{\mu \kappa } T^{\kappa }{}_{\nu \alpha } + T^{\kappa }{}_{\mu \alpha } T_{\nu }{}^{\alpha }{}_{\kappa }\big) \nonumber\\
&+ m_{T}^2 \big(T_{\nu \mu }{}^{\alpha } T_{\alpha } -  T_{\mu } T_{\nu }\big) -  \frac{2}{3} m_{t}^2 \big(2 T_{\alpha \nu }{}^{\kappa } T^{\alpha }{}_{\mu \kappa } + T^{\alpha }{}_{\mu \kappa } T^{\kappa }{}_{\nu \alpha } -  T^{\kappa }{}_{\mu \alpha } T_{\nu }{}^{\alpha }{}_{\kappa } + T_{\nu \mu }{}^{\alpha } T_{\alpha } -  T_{\mu } T_{\nu }\big) \nonumber\\
&-2 \bar{h}_{1}^{} \big(\tilde{R}_{\theta }{}^{\tau }{}_{\nu \kappa } T_{\alpha \mu }{}^{\kappa } -  \tilde{R}_{\nu \kappa }{}^{\tau }{}_{\theta } T_{\alpha \mu }{}^{\kappa } + \tilde{R}_{\theta }{}^{\tau }{}_{\mu \kappa } T_{\alpha \nu }{}^{\kappa }\big) T^{\alpha \theta }{}_{\tau } 
-2 \bar{h}_{2}^{} \big(\tilde{R}_{\kappa }{}^{\tau }{}_{\nu \theta } T_{\alpha \mu }{}^{\kappa } -  \tilde{R}_{\nu }{}^{\tau }{}_{\kappa \theta } T_{\alpha \mu }{}^{\kappa } + \tilde{R}_{\kappa }{}^{\tau }{}_{\mu \theta } T_{\alpha \nu }{}^{\kappa }\big) T^{\alpha \theta }{}_{\tau } \nonumber\\
&+ 
\bar{h}_{3}^{} \big(\tilde{R}_{\nu }{}^{\alpha \tau }{}_{\theta } T_{\alpha \mu }{}^{\kappa } T_{\kappa }{}^{\theta }{}_{\tau } + 2 \tilde{R}^{\theta \tau }{}_{\nu \kappa } T_{\alpha \mu }{}^{\kappa } T_{\tau }{}^{\alpha }{}_{\theta } + 2 \tilde{R}^{\theta \tau }{}_{\mu \kappa } T_{\alpha \nu }{}^{\kappa } T_{\tau }{}^{\alpha }{}_{\theta } -  \tilde{R}^{\tau }{}_{\theta \alpha \kappa } T_{\tau \mu }{}^{\theta } T_{\nu }{}^{\alpha \kappa }\big) \nonumber\\
&+ 
\bar{h}_{4}^{} \big(\tilde{R}_{\kappa }{}^{\tau }{}_{\nu }{}^{\theta } T_{\alpha \mu }{}^{\kappa } T_{\tau }{}^{\alpha }{}_{\theta }- \tilde{R}^{\alpha \tau }{}_{\nu \theta } T_{\alpha \mu }{}^{\kappa } T_{\kappa }{}^{\theta }{}_{\tau } -  \tilde{R}^{\alpha \tau }{}_{\mu \theta } T_{\alpha \nu }{}^{\kappa } T_{\kappa }{}^{\theta }{}_{\tau } \nonumber\\
&-  \tilde{R}_{\nu }{}^{\tau }{}_{\kappa }{}^{\theta } T_{\alpha \mu }{}^{\kappa } T_{\tau }{}^{\alpha }{}_{\theta } + \tilde{R}_{\kappa }{}^{\tau }{}_{\mu }{}^{\theta } T_{\alpha \nu }{}^{\kappa } T_{\tau }{}^{\alpha }{}_{\theta } + \tilde{R}_{\kappa }{}^{\tau }{}_{\alpha \theta } T_{\tau \mu }{}^{\theta } T_{\nu }{}^{\alpha \kappa }\big) \nonumber\\
&+ 
\bar{h}_{5}^{} \big(\tilde{R}_{\kappa }{}^{\theta }{}_{\nu }{}^{\tau } T_{\alpha \mu }{}^{\kappa } T_{\tau }{}^{\alpha }{}_{\theta }- \tilde{R}_{\nu }{}^{\tau \alpha }{}_{\theta } T_{\alpha \mu }{}^{\kappa } T_{\kappa }{}^{\theta }{}_{\tau } -  \tilde{R}_{\nu }{}^{\theta }{}_{\kappa }{}^{\tau } T_{\alpha \mu }{}^{\kappa } T_{\tau }{}^{\alpha }{}_{\theta }  \nonumber\\
&+ \tilde{R}_{\kappa }{}^{\theta }{}_{\mu }{}^{\tau } T_{\alpha \nu }{}^{\kappa } T_{\tau }{}^{\alpha }{}_{\theta } + \tilde{R}_{\kappa }{}^{\tau }{}_{\mu \theta } T_{\alpha }{}^{\theta }{}_{\tau } T_{\nu }{}^{\alpha \kappa } + \tilde{R}_{\kappa \theta \alpha }{}^{\tau } T_{\tau \mu }{}^{\theta } T_{\nu }{}^{\alpha \kappa }\big) \nonumber\\
&+ 
\bar{h}_{6}^{} \big(2 \tilde{R}_{\nu \kappa }{}^{\theta \tau } T_{\alpha \mu }{}^{\kappa } T_{\tau }{}^{\alpha }{}_{\theta }- \tilde{R}_{\theta }{}^{\tau }{}_{\nu }{}^{\alpha } T_{\alpha \mu }{}^{\kappa } T_{\kappa }{}^{\theta }{}_{\tau } -  \tilde{R}_{\theta }{}^{\tau }{}_{\mu }{}^{\alpha } T_{\alpha \nu }{}^{\kappa } T_{\kappa }{}^{\theta }{}_{\tau } -  \tilde{R}_{\theta }{}^{\tau }{}_{\mu \kappa } T_{\alpha }{}^{\theta }{}_{\tau } T_{\nu }{}^{\alpha \kappa } -  \tilde{R}_{\alpha \kappa }{}^{\tau }{}_{\theta } T_{\tau \mu }{}^{\theta } T_{\nu }{}^{\alpha \kappa }\big) \nonumber\\
&+ 
\bar{h}_{7}^{} \big[\tilde{R}_{\alpha }{}^{\kappa }{}_{\tau }{}^{\theta } T_{\kappa }{}^{\tau }{}_{\theta } T_{\nu \mu }{}^{\alpha }+ \tilde{R}_{\nu }{}^{\alpha }{}_{\theta \kappa } T_{\alpha }{}^{\kappa \theta } T_{\mu } - 2 (\tilde{R}_{\alpha }{}^{\theta }{}_{\nu \kappa } T_{\theta \mu }{}^{\kappa } + \tilde{R}_{\alpha }{}^{\theta }{}_{\mu \kappa } T_{\theta \nu }{}^{\kappa }) T^{\alpha }\big]\nonumber\\
&+ 
\bar{h}_{8}^{} \big[\tilde{R}_{\alpha \tau }{}^{\kappa \theta } T_{\kappa }{}^{\tau }{}_{\theta } T_{\nu \mu }{}^{\alpha } -  \tilde{R}_{\nu \kappa }{}^{\alpha }{}_{\theta } T_{\alpha }{}^{\kappa \theta } T_{\mu }-  \big(\tilde{R}_{\alpha \kappa \nu }{}^{\theta } T_{\theta \mu }{}^{\kappa } -  \tilde{R}_{\nu \alpha }{}^{\theta }{}_{\kappa } T_{\theta \mu }{}^{\kappa } + \tilde{R}_{\alpha \kappa \mu }{}^{\theta } T_{\theta \nu }{}^{\kappa } + \tilde{R}_{\alpha \kappa \mu \theta } T_{\nu }{}^{\kappa \theta }\big) T^{\alpha }\big] \nonumber\\
&+ 
\bar{h}_{9}^{} \big(\tilde{R}^{\kappa }{}_{\tau \alpha }{}^{\theta } T_{\kappa }{}^{\tau }{}_{\theta } T_{\nu \mu }{}^{\alpha } + \tilde{R}_{\kappa }{}^{\theta }{}_{\nu \alpha } T_{\theta \mu }{}^{\kappa } T^{\alpha } + \tilde{R}_{\nu }{}^{\theta }{}_{\alpha \kappa } T_{\theta \mu }{}^{\kappa } T^{\alpha } + \tilde{R}_{\kappa }{}^{\theta }{}_{\mu \alpha } T_{\theta \nu }{}^{\kappa } T^{\alpha } -  \tilde{R}^{\alpha }{}_{\kappa \nu \theta } T_{\alpha }{}^{\kappa \theta } T_{\mu } -  \tilde{R}^{\alpha }{}_{\kappa \mu \theta } T_{\alpha }{}^{\kappa \theta } T_{\nu }\big) \nonumber\\
&+ 
\bar{h}_{10}^{} \big(\tilde{R}_{\tau }{}^{\theta }{}_{\alpha }{}^{\kappa } T_{\kappa }{}^{\tau }{}_{\theta } T_{\nu \mu }{}^{\alpha } - 2 \tilde{R}_{\nu \kappa \alpha }{}^{\theta } T_{\theta \mu }{}^{\kappa } T^{\alpha } -  \tilde{R}_{\theta \kappa \mu \alpha } T_{\nu }{}^{\kappa \theta } T^{\alpha } + \tilde{R}_{\theta \kappa \nu }{}^{\alpha } T_{\alpha }{}^{\kappa \theta } T_{\mu } + \tilde{R}_{\theta \kappa \mu }{}^{\alpha } T_{\alpha }{}^{\kappa \theta } T_{\nu }\big) \nonumber\\
&+ 
\bar{h}_{11}^{} \big(\tilde{R}^{\theta \tau }{}_{\nu }{}^{\alpha } T_{\alpha \mu }{}^{\kappa } T_{\tau \kappa \theta } + \tilde{R}_{\nu }{}^{\alpha \theta \tau } T_{\alpha \mu }{}^{\kappa } T_{\tau \kappa \theta } + \tilde{R}^{\theta \tau }{}_{\mu }{}^{\alpha } T_{\alpha \nu }{}^{\kappa } T_{\tau \kappa \theta } \nonumber\\
& -  \tilde{R}^{\alpha }{}_{\kappa }{}^{\tau }{}_{\theta } T_{\alpha \mu }{}^{\kappa } T_{\tau \nu }{}^{\theta } -  \tilde{R}^{\tau }{}_{\theta }{}^{\alpha }{}_{\kappa } T_{\alpha \mu }{}^{\kappa } T_{\tau \nu }{}^{\theta } + \tilde{R}^{\theta \tau }{}_{\mu \kappa } T_{\tau \alpha \theta } T_{\nu }{}^{\alpha \kappa }\big) \nonumber\\
&+ 
2 \bar{h}_{12}^{} \big(\tilde{R}^{\alpha \tau }{}_{\nu }{}^{\theta } T_{\alpha \mu }{}^{\kappa } T_{\tau \kappa \theta } + \tilde{R}^{\alpha \tau }{}_{\mu }{}^{\theta } T_{\alpha \nu }{}^{\kappa } T_{\tau \kappa \theta } -  \tilde{R}^{\alpha \tau }{}_{\kappa \theta } T_{\alpha \mu }{}^{\kappa } T_{\tau \nu }{}^{\theta }\big) \nonumber\\
&+ 
\bar{h}_{13}^{} \big(\tilde{R}^{\alpha \theta }{}_{\nu }{}^{\tau } T_{\alpha \mu }{}^{\kappa } T_{\tau \kappa \theta } + \tilde{R}_{\nu }{}^{\tau \alpha \theta } T_{\alpha \mu }{}^{\kappa } T_{\tau \kappa \theta } + \tilde{R}^{\alpha \theta }{}_{\mu }{}^{\tau } T_{\alpha \nu }{}^{\kappa } T_{\tau \kappa \theta } -  \tilde{R}^{\alpha }{}_{\theta \kappa }{}^{\tau } T_{\alpha \mu }{}^{\kappa } T_{\tau \nu }{}^{\theta } \nonumber\\
& -  \tilde{R}_{\kappa }{}^{\tau \alpha }{}_{\theta } T_{\alpha \mu }{}^{\kappa } T_{\tau \nu }{}^{\theta } -  \tilde{R}_{\kappa }{}^{\tau }{}_{\mu }{}^{\theta } T_{\tau \alpha \theta } T_{\nu }{}^{\alpha \kappa }\big) \nonumber\\
&+ 
\bar{h}_{14}^{} \big[2 \tilde{R}_{\nu }{}^{\theta \alpha \tau } T_{\alpha \mu }{}^{\kappa } T_{\tau \kappa \theta } - 2 \big(\tilde{R}_{\kappa \theta }{}^{\alpha \tau } T_{\alpha \mu }{}^{\kappa } T_{\tau \nu }{}^{\theta } + \tilde{R}_{\kappa }{}^{\theta }{}_{\mu }{}^{\tau } T_{\tau \alpha \theta } T_{\nu }{}^{\alpha \kappa }\big)\big] \nonumber\\
&- 
\bar{h}_{15}^{} \big[\tilde{R}_{\nu }{}^{\tau }{}_{\mu }{}^{\alpha } T_{\alpha }{}^{\kappa \theta } T_{\tau \kappa \theta } +  \tilde{R}^{\alpha }{}_{\mu } T_{\alpha \kappa \theta } T_{\nu }{}^{\kappa \theta }+2 \tilde{R}^{\alpha \kappa } \big(T_{\alpha \nu }{}^{\theta } T_{\kappa \mu \theta } + T_{\alpha \mu }{}^{\theta } T_{\kappa \nu \theta }\big)\big]\nonumber\\
&+ 
\bar{h}_{16}^{} \big(\tilde{R}_{\nu }{}^{\tau }{}_{\mu }{}^{\alpha } T_{\alpha }{}^{\kappa }{}_{\theta } T_{\kappa }{}^{\theta }{}_{\tau } -  \tilde{R}_{\nu }{}^{\alpha } T_{\alpha }{}^{\kappa \theta } T_{\kappa \mu \theta } -  \tilde{R}^{\alpha \kappa } T_{\kappa \nu }{}^{\theta } T_{\theta \mu \alpha } -  \tilde{R}^{\alpha \kappa } T_{\kappa \mu }{}^{\theta } T_{\theta \nu \alpha } -  \tilde{R}^{\alpha \kappa } T_{\kappa \mu \theta } T_{\nu \alpha }{}^{\theta } -  \tilde{R}^{\alpha }{}_{\mu } T_{\kappa \alpha \theta } T_{\nu }{}^{\kappa \theta }\big) \nonumber\\
&+ 
\bar{h}_{17}^{} \big(\tilde{R}_{\nu }{}^{\alpha }{}_{\mu }{}^{\tau } T_{\alpha }{}^{\kappa }{}_{\theta } T_{\kappa }{}^{\theta }{}_{\tau } -  \tilde{R}^{\alpha }{}_{\nu } T_{\alpha }{}^{\kappa \theta } T_{\kappa \mu \theta } -  \tilde{R}^{\alpha }{}_{\mu } T_{\alpha }{}^{\kappa \theta } T_{\kappa \nu \theta } -  \tilde{R}^{\alpha \kappa } T_{\alpha \nu }{}^{\theta } T_{\theta \mu \kappa } -  \tilde{R}^{\alpha \kappa } T_{\alpha \mu }{}^{\theta } T_{\theta \nu \kappa } -  \tilde{R}^{\alpha \kappa } T_{\alpha \mu \theta } T_{\nu \kappa }{}^{\theta }\big) \nonumber\\
&- 
\bar{h}_{18}^{} \big( \tilde{R}_{\nu }{}^{\theta }{}_{\mu \tau } T_{\alpha }{}^{\kappa \tau } T^{\alpha }{}_{\kappa \theta } +  \tilde{R}^{\alpha }{}_{\nu } T_{\kappa \mu }{}^{\theta } T^{\kappa }{}_{\alpha \theta } +  \tilde{R}_{\nu }{}^{\alpha } T_{\kappa \mu }{}^{\theta } T^{\kappa }{}_{\alpha \theta } +  \tilde{R}^{\alpha }{}_{\mu } T_{\kappa \nu }{}^{\theta } T^{\kappa }{}_{\alpha \theta } +  \tilde{R}^{\alpha \kappa } T_{\theta \mu \kappa } T^{\theta }{}_{\nu \alpha } +  \tilde{R}^{\alpha \kappa } T_{\theta \mu \alpha } T^{\theta }{}_{\nu \kappa }\big) \nonumber\\
&+ 
\bar{h}_{19}^{} \big(\tilde{R}^{\alpha \kappa } T_{\theta \mu \kappa } T_{\nu \alpha }{}^{\theta } + \tilde{R}^{\alpha \kappa } T_{\theta \mu \alpha } T_{\nu \kappa }{}^{\theta }- \tilde{R}_{\nu }{}^{\tau }{}_{\mu \theta } T_{\alpha }{}^{\kappa \theta } T_{\kappa }{}^{\alpha }{}_{\tau } -  \tilde{R}^{\alpha }{}_{\nu } T_{\kappa \mu }{}^{\theta } T_{\theta \alpha }{}^{\kappa } -  \tilde{R}_{\nu }{}^{\alpha } T_{\kappa \mu }{}^{\theta } T_{\theta \alpha }{}^{\kappa } -  \tilde{R}^{\alpha }{}_{\mu } T_{\kappa \nu }{}^{\theta } T_{\theta \alpha }{}^{\kappa }\big) \nonumber\\
&+ 
\bar{h}_{20}^{} \big[\tilde{R}^{\alpha \kappa } \big(T_{\nu \mu \kappa } T_{\alpha } + T_{\nu \mu \alpha } T_{\kappa }\big) -  \tilde{R}_{\nu \kappa \mu \alpha } T^{\alpha } T^{\kappa } -  \tilde{R}^{\alpha }{}_{\nu } T_{\alpha } T_{\mu } -  \tilde{R}_{\nu }{}^{\alpha } T_{\alpha } T_{\mu } -  \tilde{R}^{\alpha }{}_{\mu } T_{\alpha } T_{\nu }\big] \nonumber\\
&+ 
\bar{h}_{21}^{} \big[\tilde{R}_{\nu }{}^{\kappa }{}_{\mu \theta } T_{\kappa }{}^{\alpha \theta } T_{\alpha }+ \tilde{R}^{\alpha \kappa } \big(T_{\alpha \kappa \theta } T_{\nu \mu }{}^{\theta } + T_{\alpha \nu \kappa } T_{\mu } + T_{\alpha \mu \kappa } T_{\nu }\big)-  \big(\tilde{R}^{\alpha }{}_{\nu } T_{\alpha \mu }{}^{\kappa } + \tilde{R}^{\alpha }{}_{\mu } T_{\alpha \nu }{}^{\kappa }\big) T_{\kappa }\big]\nonumber\\
& + 
\bar{h}_{22}^{} \big[\tilde{R}_{\nu \theta \mu }{}^{\kappa } T_{\kappa }{}^{\alpha \theta } T_{\alpha } + \tilde{R}^{\alpha \kappa } \big(T_{\kappa \alpha \theta } T_{\nu \mu }{}^{\theta } + T_{\kappa \nu \alpha } T_{\mu } + T_{\kappa \mu \alpha } T_{\nu }\big)-  \big(\tilde{R}_{\nu }{}^{\alpha } T_{\alpha \mu }{}^{\kappa } + \tilde{R}^{\alpha }{}_{\mu } T_{\nu \alpha }{}^{\kappa }\big) T_{\kappa }\big] \nonumber\\
&+ 
\bar{h}_{23}^{} \big[\tilde{R}_{\nu \theta \mu \kappa } T_{\alpha }{}^{\kappa \theta } T^{\alpha } + \big(\tilde{R}^{\alpha }{}_{\nu } T_{\kappa \mu \alpha } -  \tilde{R}_{\nu }{}^{\alpha } T_{\kappa \mu \alpha } + \tilde{R}^{\alpha }{}_{\mu } T_{\kappa \nu \alpha }\big) T^{\kappa } + \tilde{R}^{\alpha \kappa } \big(T_{\theta \alpha \kappa } T_{\nu \mu }{}^{\theta } -  T_{\nu \alpha \kappa } T_{\mu }\big)\big] \nonumber\\
&
-2 \bar{h}_{24}^{} \big(\tilde{R}_{\nu \mu } T_{\alpha }{}^{\kappa \theta } T^{\alpha }{}_{\kappa \theta } + 2 \tilde{R} T_{\alpha \mu }{}^{\kappa } T^{\alpha }{}_{\nu \kappa }\big) 
-2 \bar{h}_{25}^{} \big(\tilde{R}_{\nu \mu } T_{\alpha }{}^{\kappa \theta } T_{\kappa }{}^{\alpha }{}_{\theta } + \tilde{R} T_{\alpha \mu }{}^{\kappa } T_{\kappa \nu }{}^{\alpha } + \tilde{R} T_{\alpha \mu \kappa } T_{\nu }{}^{\alpha \kappa }\big)\nonumber\\
& + 
\bar{h}_{26}^{} \big[2 \tilde{R} T_{\nu \mu }{}^{\alpha } T_{\alpha } - 2 \big(\tilde{R}_{\nu \mu } T_{\alpha } T^{\alpha } + \tilde{R} T_{\mu } T_{\nu }\big)\big]\,.
\end{align}

\section{Lagrangian coefficients of the Poincaré cubic action with stable vector and axial sectors}\label{appe2}

The ghostly instabilities present in the vector and axial sectors of quadratic PG theory can be eliminated by including cubic order invariants defined from mixing terms of the curvature and torsion tensors~\cite{Bahamonde:2024sqo}. Thus, the Lagrangian coefficients of the gravitational action~\eqref{cubicPGmodel_fulltorsion}, which contains all the possible mixing terms of cubic order, are restricted by the corresponding stability conditions as
\begin{equation}
    c_{2} = 2c_{1}\,, \quad h_{2}=-\,\frac{h_{1}}{2}\,, \quad h_{3}= -\,\frac{1}{6}\left(c_{1}+6h_{13}\right),\quad h_{4}= \frac{h_{13}}{2}\,,\quad h_{14}= -\,2h_{13}\,,\quad h_{15}=4h_{13}\,.
\end{equation}

Then, the resulting action can be equivalently expressed as Eq.~\eqref{cubicPGmodel_irreducible_torsion}, in virtue of the parametrisation
\begin{align}
    \bar{h}_1&=\frac{1}{3} \left(h_5-h_7+6 h_{20}\right),\quad \bar{h}_2=\frac{1}{3} \left[h_6-h_8+6 \left(h_{24}-2 h_{20}\right)\right],\quad\bar{h}_3=\frac{2}{3} \left(h_5-3 h_{21}\right),\\
    \bar{h}_4&=\frac{1}{3} \left[2 h_6-h_8+2 \left(h_9+6 h_{20}+6 h_{22}-3 h_{23}-3 h_{24}\right)\right],\quad\bar{h}_5=2 h_{24}-\frac{2 h_6}{3}-\frac{h_8}{3}-4 h_{20}-4 h_{21},\\
    \bar{h}_6&=\frac{2}{3} \left[h_5-3 \left(2 h_{20}+h_{22}\right)\right],\quad\bar{h}_7=-\,\frac{2}{9} \left(2 h_5+h_6-h_7-h_8+18 h_{13}+3 h_{17}+12 h_{20}-9 h_{21}-3 h_{24}\right),\\
    \bar{h}_8&=\frac{2}{9} \left(h_7-2 h_5-h_6+h_8+36 h_{13}-3 h_{17}+24 h_{20}-18 h_{21}-6 h_{24}\right),\\
    \bar{h}_9&=\frac{1}{9}
   \left(4 h_5+2 h_6-2 h_7-2 h_9-3 h_{16}-24 h_{20}+36 h_{22}-12 h_{23}+12 h_{24}\right),\\
   \bar{h}_{10}&=\frac{1}{9} \left(2 h_7-4 h_5-2 h_6+2 h_9+3 h_{16}-12 h_{20}+18 h_{22}-6 h_{23}+6 h_{24}\right),\quad\bar{h}_{11}=4 \left(h_{21}+h_{22}\right)-\frac{2 h_7}{3},\\
   \bar{h}_{12}&=\frac{1}{3} \left[h_9-h_8+6 \left(h_{23}-2 h_{22}\right)\right],\quad\bar{h}_{13}=\frac{1}{3} \left[h_8+2 \left(h_9+6 h_{21}+6 h_{22}-3 h_{23}\right)\right],\quad  \bar{h}_{14}=-\,4 h_{21},\\
  \bar{h}_{15}&=\frac{1}{9} \left(2 h_5-3 c_1+h_6+2 h_7+h_8+h_9-2 h_{10}+h_{11}-18 h_{13}-6 h_{23}-6 h_{24}+6 h_{25}\right),\\
  \bar{h}_{16}&=\frac{1}{9} \left(6 c_1-4 h_5-2 h_6-4 h_7-2 h_8-2
   h_9-2 h_{10}+h_{11}+36 h_{13}+12 h_{23}+12 h_{24}-12 h_{25}-36 h_{26}\right),\\
   \bar{h}_{17}&=\frac{1}{9} \left(6 c_1-4 h_5-2 h_6-4 h_7-2 h_8-2 h_9-2 h_{10}+h_{11}+36 h_{13}+12 h_{23}+12 h_{24}+6 h_{25}+36 h_{26}\right),\\
   \bar{h}_{18}&=\frac{1}{9} \left(4 h_5-6 c_1+2 h_6+4 h_7+2 h_8+2 h_9-h_{10}+5 h_{11}-36 h_{13}-12 h_{23}-12 h_{24}-6 h_{25}\right),\\
   \bar{h}_{19}&=\frac{1}{9} \left(6 c_1-4 h_5-2 h_6-4 h_7-2 h_8-2 h_9+h_{10}+4 h_{11}+36 h_{13}+12 h_{23}+12 h_{24}+6 h_{25}\right),\\
   \bar{h}_{20}&=\frac{1}{9} \left(9 h_1+4 h_5+2 h_6-2 h_7-h_8-2 h_9+h_{10}-4 h_{11}-3 h_{16}+6 h_{17}-3 h_{19}\right),\\
   \bar{h}_{21}&=\frac{1}{9}
   \left[4 h_9-h_8+3 \left(h_{11}-h_{10}+24 h_{13}-h_{18}+2 h_{19}+4 h_{21}-4 h_{22}+2 h_{23}+2 h_{25}+8 h_{26}\right)\right],\\
   \bar{h}_{22}&=\frac{1}{9} \left(h_8+2 h_9-3 \left(h_{10}-h_{11}+24 h_{13}-h_{18}-h_{19}+4 h_{21}-4 h_{22}+2 h_{23}+2 h_{25}+8 h_{26}\right)\right),\\
   \bar{h}_{23}&=\frac{1}{9} \left[2 h_8-2 h_9+3 \left(24 h_{13}+2 h_{18}-h_{19}+4 h_{21}-4 h_{22}+2 h_{23}+2 h_{25}+8 h_{26}\right)\right],\\
   \bar{h}_{24}&=\frac{1}{18} \left(6 c_1-2 h_5-h_6-2 h_7-h_8-h_9+12 h_{12}+18 h_{13}+6 h_{23}+6 h_{24}\right),\\
   \bar{h}_{25}&=\frac{1}{9} \left(2 h_5-6 c_1+h_6+2 h_7+h_8+h_9+6 h_{12}-18 h_{13}-6 h_{23}-6
   h_{24}\right),\\
   \bar{h}_{26}&=\frac{1}{18} \left(2h_9-9h_1-2h_{10}+2h_{11}-12 h_{12}+6 h_{19}\right).
\end{align}

\bibliographystyle{utphys}
\bibliography{references}

\end{document}